\documentclass[onecolumn,aps,preprint,groupedaddress,showpacs,floatfix,notitlepage,color]{revtex4-2}
\usepackage{graphicx}
\usepackage{epsfig}
\usepackage{amsfonts}
\usepackage{color}
\usepackage{ulem}
\usepackage{amsmath}
\usepackage{mathrsfs}
\usepackage[autostyle]{csquotes}
\usepackage{url}
\usepackage{float}
\usepackage{siunitx}
\usepackage{subcaption}
\definecolor{purple}{rgb}{0.5, 0.0, 0.5}
\definecolor{green4}{rgb}{0.13,0.55,0.13}
\definecolor{orange}{rgb}{1.0, 0.5, 0.0}

\begin{document}

\title{Predicting the Progression of Cancerous Tumors in Mice: \\ A Machine and Deep Learning Intuition}

\author{${}^{*}$Amit K Chattopadhyay and Aimee Pascaline N Unkundiye}
\affiliation{                    
Department of Applied Mathematics and Data Science, Aston Centre for Artificial Intelligence Research and Applications (ACAIRA), Aston University, Aston Triangle, Birmingham B4 7ET, United Kingdom}
\email{a.k.chattopadhyay@aston.ac.uk}
\author{Gillian Pearce}
\affiliation{College of Engineering and Physical Sciences, Aston University, Aston Triangle, Birmingham B4 7ET, United Kingdom}
\author{Steven T Russell}
\affiliation{                    
College of Life and Health Sciences, Aston University, Aston Triangle, Birmingham B4 7ET, United Kingdom}

\begin{abstract}
\textbf{Outline}: {\color{black}{The study explores Artificial Intelligence (AI) powered modeling to predict the evolution of cancer tumor cells in mice under different forms of treatment. The AI models are analyzed  against varying ambient and systemic parameters, e.g. drug  dosage, volume of the cancer cell mass, and time taken to destroy the cancer cell mass.}} The data required for the analysis have been synthetically extracted from plots available in both published and unpublished literature (primarily using a Matlab architecture called \enquote{Grabit}), that are then statistically standardized around the same baseline for comparison. Three forms of treatment are considered - saline (multiple concentrations used), magnetic nanoparticles (mNPs) and fluorodeoxyglycose iron oxide magnetic nanoparticles (mNP-FDGs) - analyzed using three Machine Learning (ML) algorithms, Decision Tree (DT), Random Forest (RF), Multilinear Regression (MLR), and a Deep Learning (DL) module, the Adaptive Neural Network (ANN). The AI models are trained on 60-80\% data, the rest used for validation. {\color{black}{Assessed over all three forms of treatment, ANN consistently outperforms other predictive models}}. Our models predict mNP-FDG as the most potent treatment regime that kills the cancerous tumor completely in ca 13 days from the start of treatment. {\color{black}{The prediction aligns with  independent research that demonstrate the use of iron mNPs combined with hyperthermia leading to 90\% shrinkage of the tumor within 12 days, but is most likely accompanied by thermal damage to the tissues surrounding the tumor.}} The prediction is consistent across all five data standardization protocols that reconfirms the diagnostic power of the AI toolset.
\\
\textbf{Main Limitations}: The specific form of laboratory treatment (saline and mNP-FDG) can be subjective of patients and their characteristics. The use of cell lines from mice bearing MAC-16 tumor grafts in the study may not accurately replicate various tumor types, potentially impacting the accuracy of suggested 
dosages. However, the AI toolset is generic and can accommodate other treatment forms, should data be available. \\
\textbf{Objective:} {\color{black}{Comparing the most potent treatment regime for cancer tumors from two choices - saline solutions at different concentrations and mNP-FDG. The main aim is to predict a timeline for the complete removal of cancer tumors. Towards this, we employ a combination of machine, deep learning and mathematical models to compare treatments against timelines through a non-invasive personalized regime of diagnostics.}} 
\end{abstract}

\maketitle

\newpage
\section{Introduction}
In 2022 alone, 18 million people world-wide were known to have cancer, and these numbers are predicted  to double by 2040 \cite{IARC2022}. 
Traditional treatments of cancer include chemotherapy, radiotherapy and more recently immunotherapy.  However, it is known that not all cancers are equally responsive to these traditional treatments. Although, chemotherapy has been a long-standing approach to cancer, it is known to have massive side effects which can be debilitating \cite{altun2018, anand2022}. Furthermore it is known that resistance to chemotherapy drugs can develop over a period of time. Also, not all cancers are responsive to radiotherapy, and such treatment can effectively destroy healthy cells, potentially leading to a plethora of other unwanted side effects \cite{suciu2021} and \cite{rex2006}. In other words, both chemo and radiotherapy are invasive, cancerous form specific, and not necessarily palliative to patients with acute comorbidity. 

\par
Studies have shown that tumor growth can be reduced or destroyed by the action of salt solutions at varying concentrations \cite{qi2022,gao2015,huffington2014,wang2021}. However, where chemotherapy agents and new treatments are involved, such research requires the involvement of animals in experimental work, that can be an ethical challenge. This is where recent advancements in the mathematics of Artificial Intelligence (AI), combining tools from Machine and Deep Learning (ML/DL), have helped in changing the perspective. We can now use DL and ML methods not only to model the destruction of cancer cells analysing existing data from (published and unpublished) literature, but also to predict the nature and volume of containment of the cancerous tumors, even predicting the treatment success timelines. While not a substitute of laboratory based science, AI tools can complement traditional laboratory based diagnostics, thus massively reducing the need to involve animals, reduce the costs involved, and accelerating treatment timelines.

\par
Nanoparticles, including magnetic nanoparticles  (mNPs), have long been explored as {\color{black}{drug delivering agents}} to destroy cancer. The hope is that such nanoparticles would better target cancer cells thus improving treatment efficacy. {\color{black}Use of iron based magnetic  nanoparticles (mNPs) has been undertaken both in-vitro and in-vivo by \cite{subramanian2016} and \cite{Akin2018}. The in-vitro work by Subramanian, et al \cite{subramanian2016} demonstrated that using mNP-FDG without hyperthermia destroyed 72\% of the cancer cells while with hyperthermia, the number destroyed 89\%}. 
Use of Saline solutions and mNPs as treatment procedures in-vivo have been undertaken by \cite{barbosa2022,curcio2023,Ji2022,subramanian2016,suciu2021,gao2015} and \cite{Akin2018,bossmann2022,Land2019}. High saline concentration is known to restrict tumor volume growth compared to an untreated tumor or uncoated mNPs \cite{Land2019}, but is also proven to be less effective compared to iron oxide coated mNPs. When  saline solution is plasma activated, it can selectively induce cell death in tumors while sparing most healthy cells \cite{gao2015}. {\color{black}{Reactive species like ions, electrons, and neutral
particles in Plasma Activated Saline (PAS) have been used in laboratory based experiments on
tumors. However they can degrade over time, thus reducing their effectiveness.
Treating tumors with PAS is also not a precision targeting technique, and there are problems
that arise concerning uniform distribution within the tumors, which do not arise when using magnetic nanoparticles (mNPs)
(which are best in precision targeting).}} However, the abilities of PAS and saline solutions to shrink or control tumor volume growth compared to mNPs proves that  mNPs on their own are not as  reliable or safe to use alone. This explains attempts at enhancing the effectiveness of mNPs by combining them with various other substances. The choice of these add-on materials are guided by a high surface to volume ratio, preventing the interaction of the mNP surface within the body during in-vivo applications. This provides them with better stability, sustainability and strength \cite{malhotra2020,pankhurst2003}. Some of these studies by \cite{dibona2015,suciu2021} raised concerns regarding the retention of mNPs in the body which could lead to liver damage. For instance, \cite{curcio2023} investigated the retention of liposomes and magnetosomes mNPs in tumors through intravenous injection. Their research highlights that only 2.27\% magnetic liposomes and magnetosomes combined reach the tumor site. \cite{barbosa2022,malhotra2020} addressed these concerns, suggesting surface modification, time and dosage adjustments to solve such issues.

\par 
The present study pitchforks Iron oxide-based nanoparticles ${\text{Fe}}_3{\text{O}}_4$ as the preferred option due to their superparamagnetic behavior, biocompatibility, and chemical stability. Also of importance is their amphoteric nature,the feature of varying surface charge based on the ambient pH \cite{khedri2018,montiel2022}. ${\text{Fe}}_3{\text{O}}_4$ is currently in clinical use as MRI contrast agents \cite{vallabani2018}. The coating composition and thickness influence the degradation and colloidal stability in the human body \cite{dadfar2019}. 
Uncoated core ${\text{Fe}}_3{\text{O}}_4$ mNPs are highly prone to aggregate and oxidise in air, resulting in a significant reduction in their magnetism and dispersion. Therefore, it is crucial to consider the functionality of ${\text{Fe}}_3{\text{O}}_4$ mNPs with better coatings\cite{shen2018}. 
Fluorodeoxyglucose (FDG) is a chemical compound used in the imaging of breast cancer Positron Emission Tomography (PET). It aids diagnosis due to its rapid  uptake by cancer cells making it a safe and reliable compound to use. The combination of FDG and ${\text{Fe}}_3{\text{O}}_4$ nanoparticles as  mNP-FDGs demonstrated by \cite{Akin2018, Land2019} shows promising results. Their findings indicate that mNP-FDGs are non-toxic when administered both intravenously and intratumorally at certain concentrations.

\par 
Our primary objective is to develop a reliable AI toolbox, combining Machine and Deep Learning Models, that can use data from existing literature and predict results that match independent experimental results from other groups (for similar ambient and systemic constraints). Towards the multifarious treatment forms, we have chosen Saline solutions because it has been extensively used as an intermediate metric in various studies and evaluated by several authors, e.g. \cite{Akin2018} found it to be a very effective in inhibiting cancer cell growth. {\color{black}{Both the research by Land, et al \cite{Land2019}, Gao, et al \cite{gao2015} and Qi, et al \cite{qi2022}, support these findings. Land, et al \cite{Land2019} conducted experiments on a MAC-16 tumor bearing mouse over 5 days while Gao, et al's studies were conducted 16 days, and Qi, et al \cite{qi2022} for 30 days. The datasets from Qi, et al and Gao, et al were used to validate the unpublished literature by Land, et al's findings, allowing for predictive modeling across the mNP-FDG dataset.}}

\section{Background}
It is well known that cancer cells have an increased need for glucose on account of their higher metabolic activity \cite{liberti2016}. This equally applies to an analogue compound of glucose called mNP-FDG (iron based magnetic nanoparticles conjugated with  fluorodeoxyglucose).
Cancer cells are rapidly taken up by mNP-FDG. Injected intratumorally, the mNP-FDG begins to destroy cancer cells within 10-20 minutes. 
Beyond a certain threshold concentration, the mNP-FDG is toxic to cancer cells, as shown in recent studies \cite{subramanian2016, watkins2018,unak2021}. mNP-FDG has also been shown effective against prostate cancer in-vivo in mice \cite{subramanian2016}.
 
In order to demonstrate that in itself mNP-FDG is not a toxic agent, tail veins of mice were injected with a concentration of mNP-FDG that was twice the concentration that would be used in human scaled down for body volume and body mass. The mice showed no detrimental effects and there was no mNP-FDG apparent in their tissue at 3 and 6 months following injection by Watkins, et al \cite{watkins2018}.

While not a systemic toxic agent itself, mNP-FDG can have several therapeutic advantages. First, it does not apparently have any of the unwanted side effects of chemotherapy (hair loss, nausea, etc.) or  radiotherapy. Also, it can be implemented against all known cancer types and importantly, cancers cannot develop resistance over time to mNP-FDG as they can with chemotherapy cancer drugs. What we are trying to analyze here is to quantify the impact of mNP-FDG in containing cancerous tumors, and predicting the impact timelines of such a treatment procedure.

\section{Methodology}
This section outlines details of the data extracted from both published and unpublished literature using software tools, the standardization protocols used, and the AI-powered analytical methods employed for this study.

\subsection{Materials/Dataset}
Synthetic datasets outlining changes in the volumes of cancerous tumors of mice suffering from cancer have been studied, in the presence and absence of treatments (saline, mNP, mNP-FDG) \cite{Land2019}. The data  were extracted from the plots from this dissertation work and other studies \cite{qi2022}, spanning different time periods, using a Matlab-tool called \enquote{Grabit}. This tool can generate data from any given plots which are experimentally measured by others but once Grabit-extracted, these are synthetic data for us, allaying ethical concerns. Approximately 600 data points are captured for each sector (untreated, saline treated, mNP treated, mNP-FDG treated and PAS treated) using Grabit. {\color{black}{We divide the total dataset of cardinality $n = 600$ into two sectors, a training set $n_1 =$ 80\% data, and a test set $n_2 =$ 20\%, such that $n=n_1+n_2$.  Although real experimental dataset would typically contain impurities such as noise, missing values, and outliers, our Grabit-extracted dataset do not require cleaning, ensuring data completeness. To ensure uniformity, for prediction, the data from different sources are appropriately scaled without any loss of generality.}}

\subsection{Methods}
Data are obtained from multiple sources \cite{Land2019,gao2015,qi2022}. The first key step is to set a flat baseline across all different datasets, through statistical standardization, to ensure uniformity in statistics. Five standardisation protocols are used - Mean Standardisation, Min-Max Standardisation, Unit vector Standardisation, Box Cox Standardisation and Logarithmic Amplitude adjustment. One of each compared datasets is re-scaled and mapped using nonlinear least squares, incorporating {\it shifting} and {\it scaling} factors ranging from -0.2 to +1.25 to ensure best statistical fitting.
{\color{black}{Our study clearly points to two of these standardizarion methods for best fit:
\textbf{Min-Max Standardization} and \textbf{Mean Standardization}.}} \textbf{Min-Max Standardization} transforms each data point \( v \)  to \( v' \) over the range [0,1] as shown in Figures \ref{fig5} and \ref{fig6}:

\[ v' = \frac{v - x_{\text{min}}}{x_{\text{max}} - x_{\text{min}}} \]

\noindent
where \( x_{\text{min}} \) is the minimum volume that the data is shifted to while \( x_{\text{max}} \) is the maximum volume  of the dataset. The scaling factor here is $(x_{\text{max}}-x_{\text{min}})$.
 
On the other hand, the \textbf{Mean Standardization} transforms each data point \( v \)  to \( v' \), scaling them using standard deviation \( \sigma_x \) of the dataset and shifted by the mean $\bar{x}$:

\[ v' = \frac{v - \bar{x}}{\sigma_x} \],

where the data $v$ is shifted over mean $\bar{x}$ and scaled by the standard deviation $\sigma_x$. 
Both {\it Min-Max} and {\it Mean} standardization protocols evidence better alignment of the data and hence have been consistently used in our follow-up modelling. However, this, does not account for different ambient conditions under which these laboratory experiments were originally undertaken. 

\par
{\color{black}{Four independent AI models have been used for analysis. The last 25 days of saline data from Qi, et al \cite{qi2022} are used to predict tumor volume for the first 5 days, comparing against the 5 days of data from Land, et al \cite{Land2019}. The Gao, et al \cite{gao2015} data is similarly used: last 11 days from its 16-day study period are used for training the models that are then compared against the Land, et al data over its 5-day period. Both data sets, Qi, et al and Gao, et al are also compared against their own first 5 days'predictions profile, after training over the remainder of their respective periods.}}

\subsection{Discussion and Results}
\begin{figure}[H]
     \setkeys{Gin}{width=1.2\linewidth,height=0.32\textheight}
 \begin{subfigure}{0.45\textwidth}
    \includegraphics[width=1\linewidth]{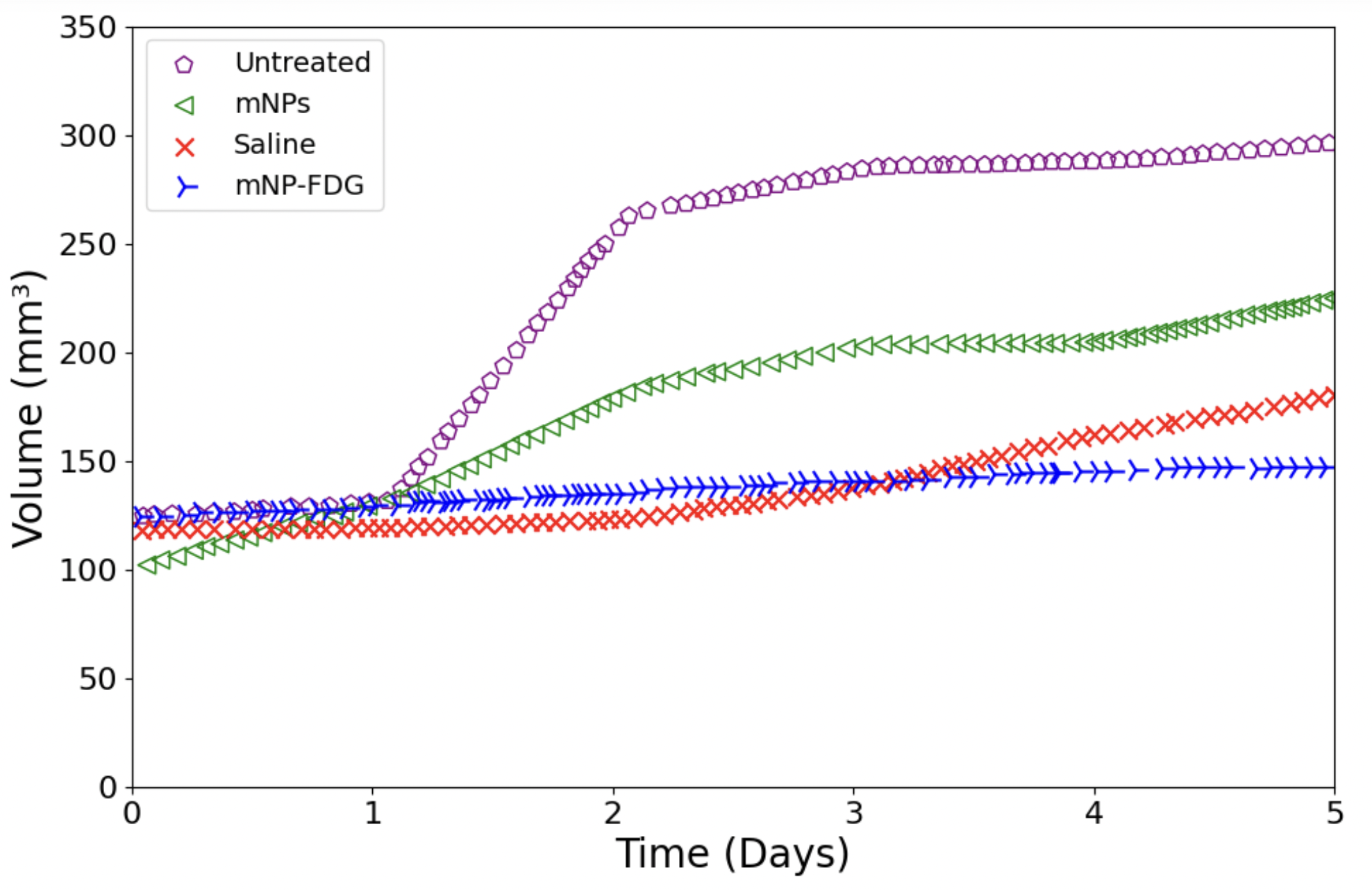}
    \caption{{\color{black}{Matlab-Grabit extracted data from Land, et al: Time evolution (5 days) of tumor volumes in mice using various forms of treatment \cite{Land2019}.}}}
    \label{fig1a}
\end{subfigure}
\hfill
\begin{subfigure}{0.45\textwidth}
    \includegraphics[width=1.2\linewidth, height=0.32\textheight]{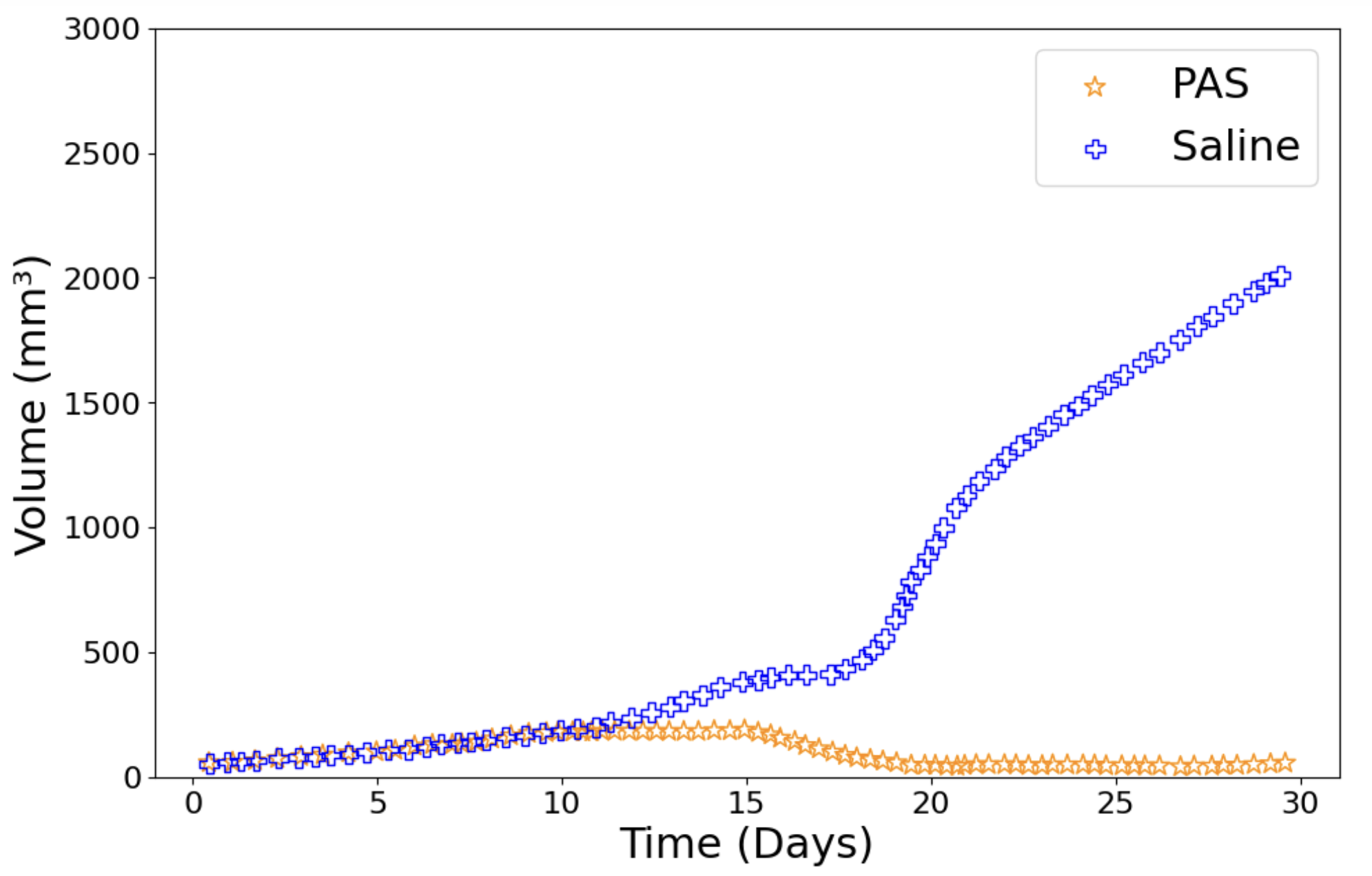}
    \caption{{\color{black}{Matlab-Grabit extracted data from Qi, et al: Time evolution (30 days) of tumor volumes in mice using various forms of treatment \cite{qi2022}. }}}
    \label{fig1b}
    \end{subfigure}
    \medskip
 \begin{subfigure}{0.49\textwidth}
     \includegraphics[width=\textwidth]{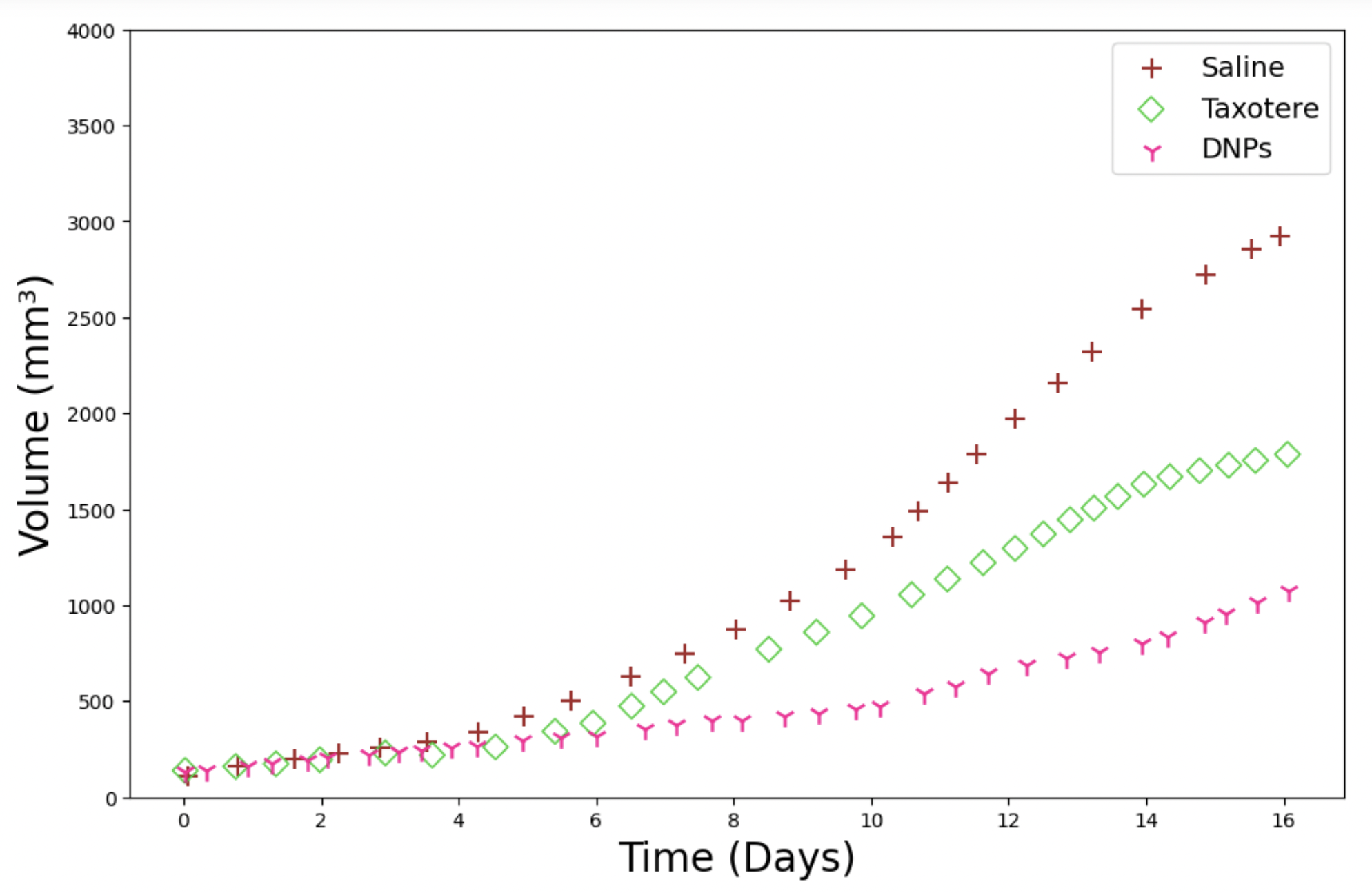}
    \caption{{\color{black}{Matlab-Grabit extracted data from Gao, et al: Time evolution (16 days) of tumor volumes in mice using various forms of treatment \cite{gao2015}.}}}
    \label{fig1c}
\end{subfigure}
\hfill
\begin{subfigure}{0.40\textwidth}
    \includegraphics[width=1.2\linewidth]{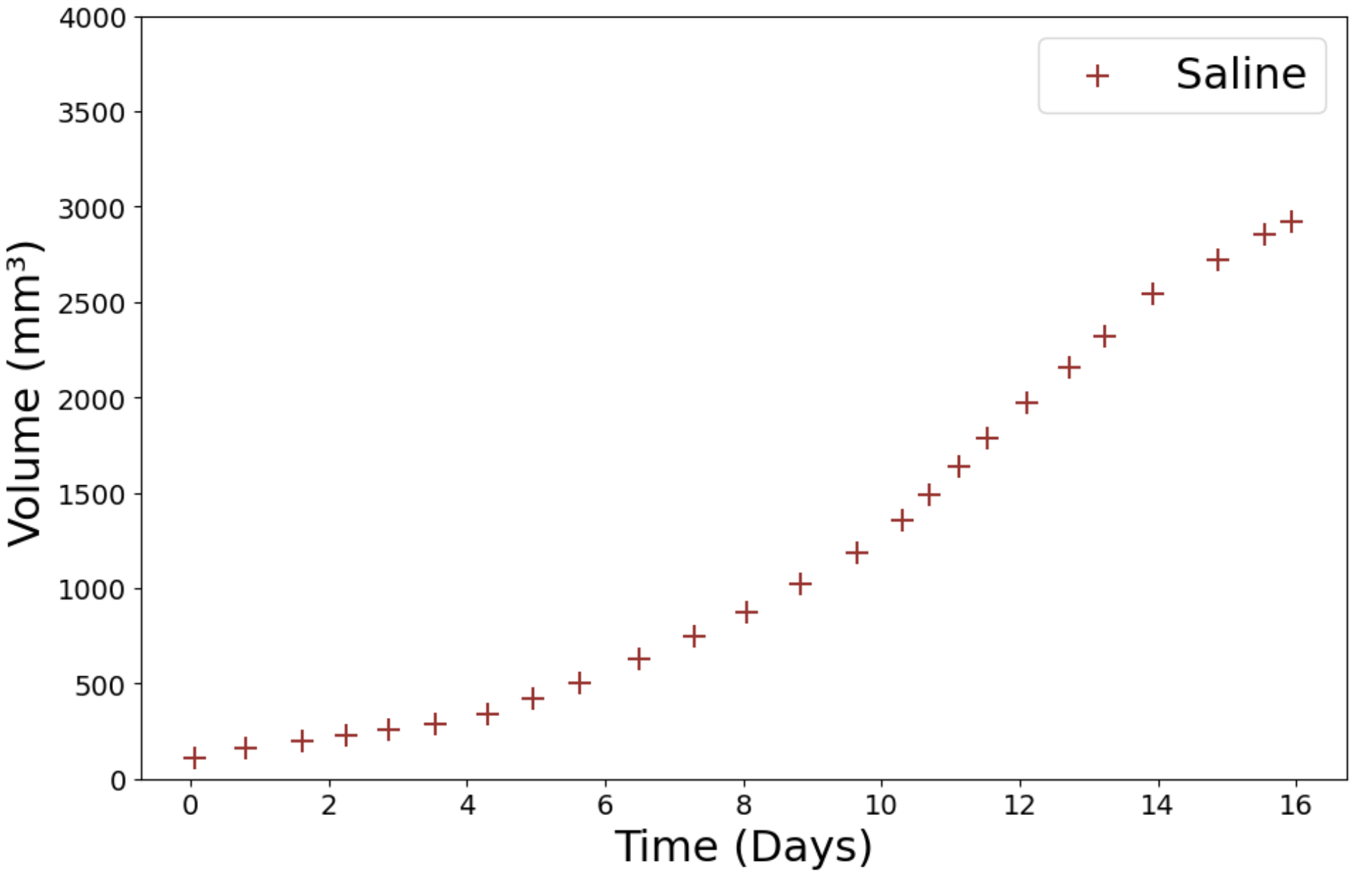}
    \caption{\color{black}{Matlab-Grabit data: Time evolution of tumor volumes in mice using Saline treatment from Gao, et al \cite{gao2015}.}}
    \label{fig:1d}
    \end{subfigure}
\caption{{\color{black}{Matlab Extracted data representing the time evolution of cancerous tumors under various treatment regimes \cite{qi2022,gao2015,Land2019}.}} }
\label{fig1}
\end{figure}

{\color{black} Figure {\ref{fig1a} displays data representing the progression of cancerous tumors in MAC-16 mice that are separately treated with saline, mNP, mNP-FDG solutions, and an untreated control group. The data have been extracted from Land et al's dissertation covering a 5-day period \cite{Land2019} using Matlab-Grabit software. Additionally, Figures \ref{fig1b} and \ref{fig1c} display broader datasets from other independent studies by Qi, et ali \cite{qi2022} and Gao, et al \cite{gao2015}, involving saline solutions and other treatments over 30 days and 16 days respectively. Multiple data sources are used to validate the findings from the unpublished study by Land, et al \cite{Land2019}}}.

\subsubsection{Comparative Tumor Evolution}
Multiple data sources, experimented under different ambient conditions, are compared in this section.

\begin{figure}[H]
   \setkeys{Gin}{width=1.2\linewidth,height=0.32\textheight}
 \begin{subfigure}{0.45\textwidth}
    \includegraphics[width=1\linewidth]{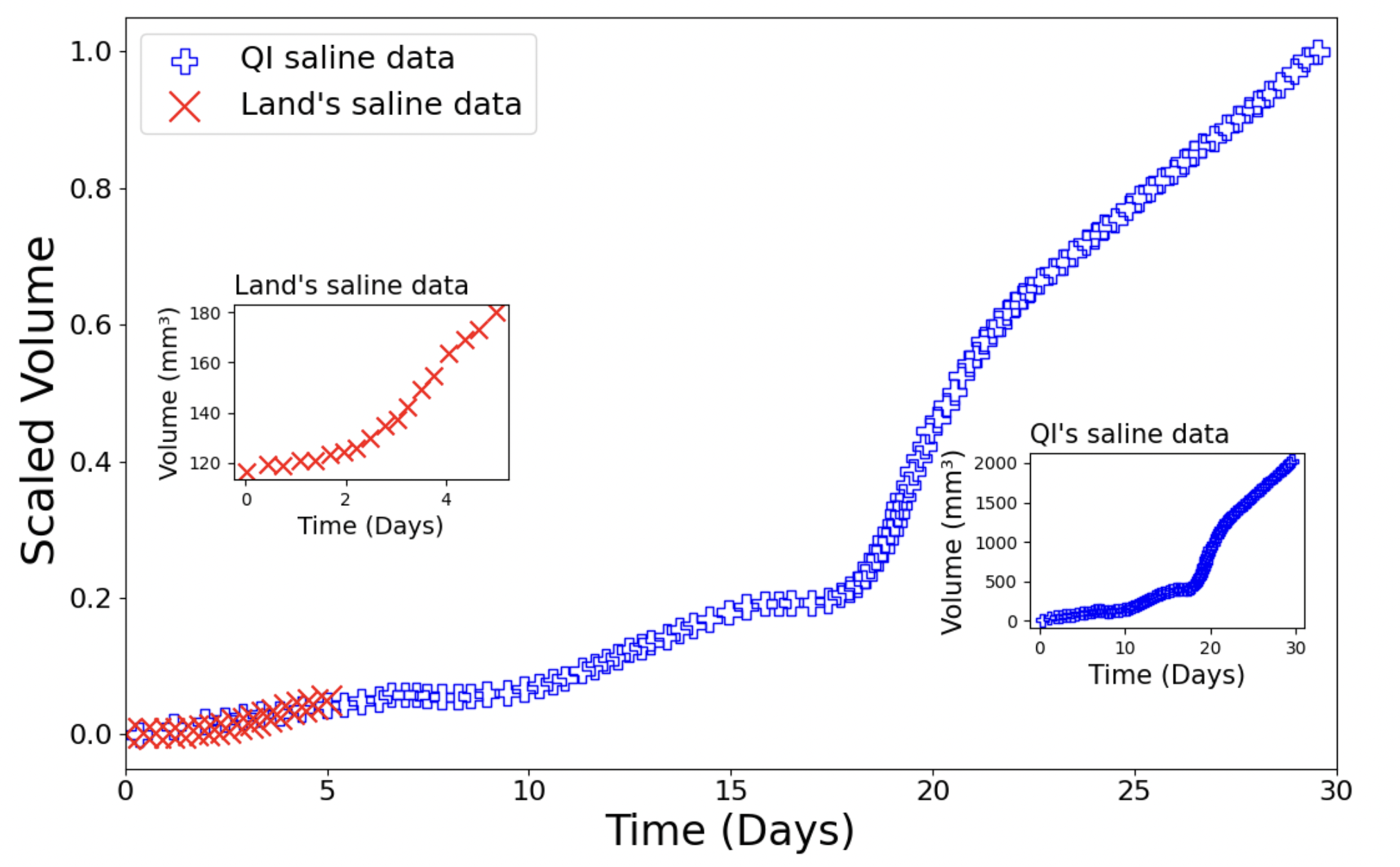}
    \caption{{\color{black}{Min-Max scaled 5-days' data from Land, et al \cite{Land2019} overlaid on scaled 30-days' from Qi, et al \cite{qi2022}.}}}
    \label{fig2a}
    \end{subfigure}
 \hfill
 \begin{subfigure}{0.45\textwidth}
    \includegraphics[width=1\linewidth]{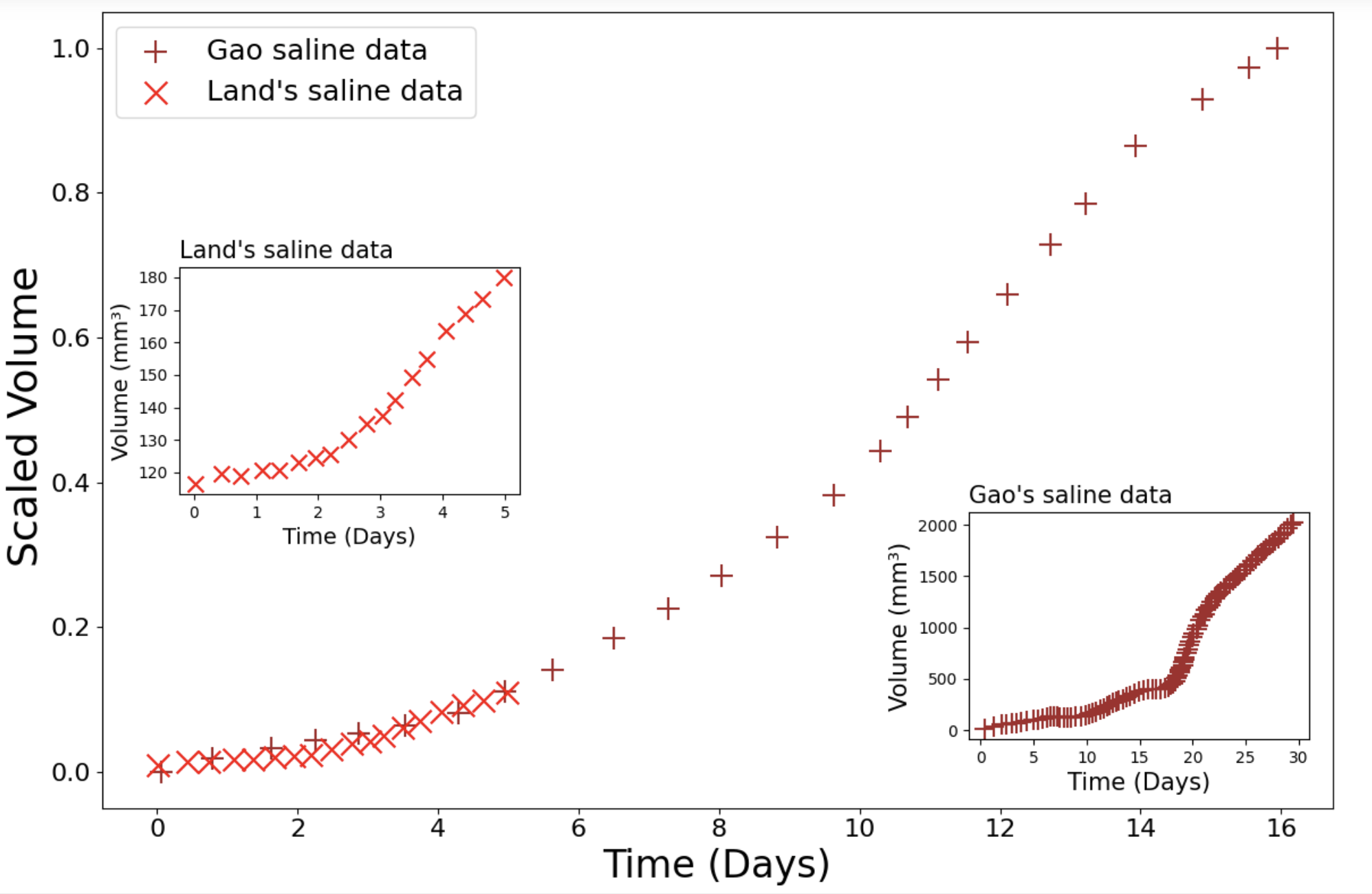}
    \caption{{{\color{black}{Min-Max scaled 5-days' data from Land, et al \cite{Land2019} overlaid on scaled 16-days' from Gao, et al \cite{gao2015}.}}}}
     \label{fig2b}
    \end{subfigure}
\caption{ {\color{black} 5-days' data from Land, et al \cite{Land2019} overlaid on scaled data from Qi, et al \cite{qi2022} and Gao, et al \cite{gao2015}, with insets for broader comparative analysis. The top inset shows saline-treated data from Land, et al while the bottom insets display data from Qi, et al and Gao, et al respectively.}}
\label{fig2}
\end{figure}

\subsubsection{Functional Analysis from Support Vector Regression (SVR)}

\begin{figure}[H]
 \setkeys{Gin}{width=1.2\linewidth,height=0.32\textheight}
 \begin{subfigure}{0.49\textwidth}
    \includegraphics[width=1\linewidth]{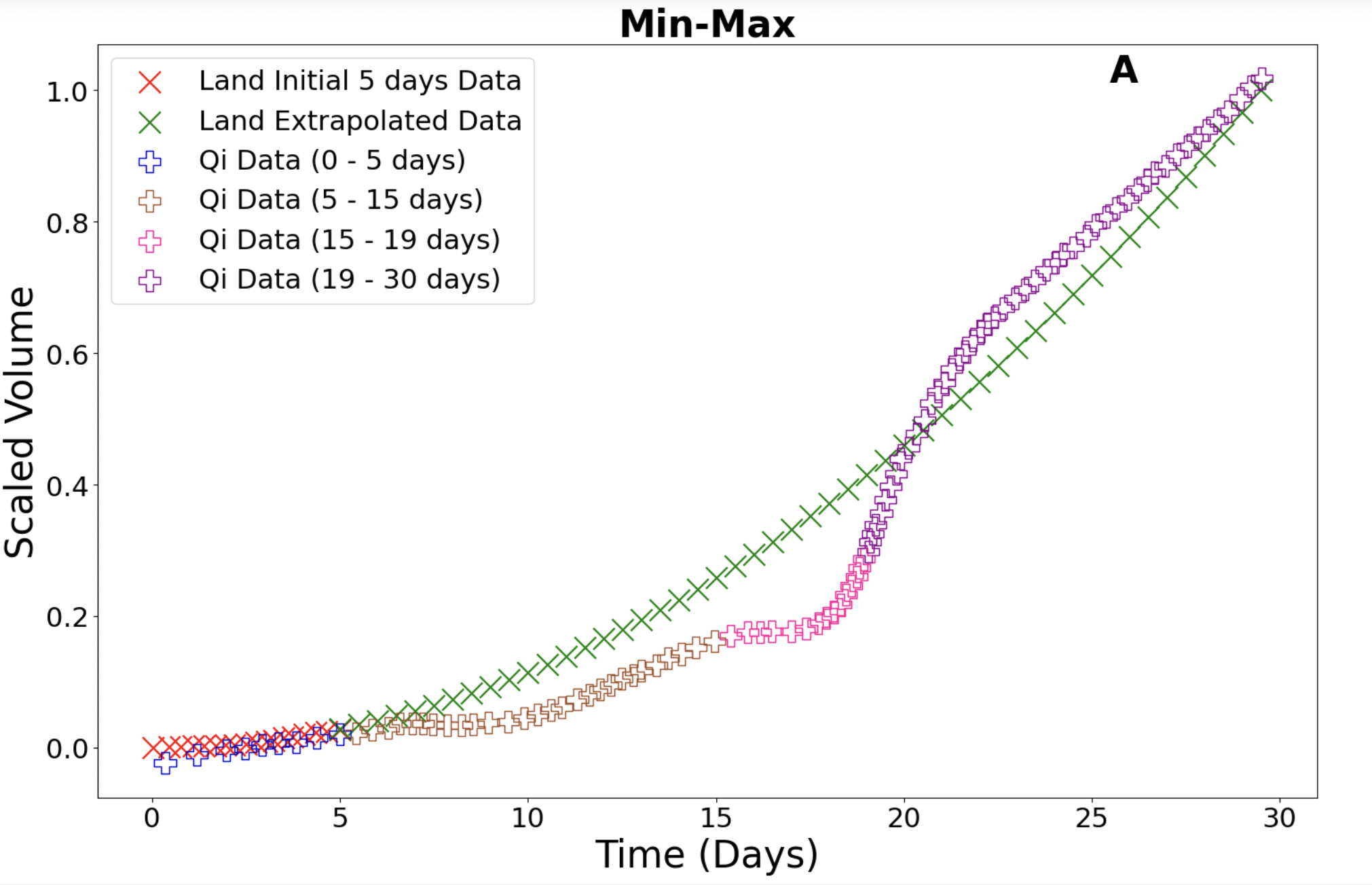}
    \caption{{\color{black}{Min-Max Standardization with respective shifting and scaling factors of 0.03 and 1.2, comparing Land, et al \cite{Land2019} versus Qi, et al \cite{qi2022} data.}}}
    \label{fig3a}
\end{subfigure}
 \hfill
 \begin{subfigure}{0.49\textwidth}
    \includegraphics[width=1\linewidth]{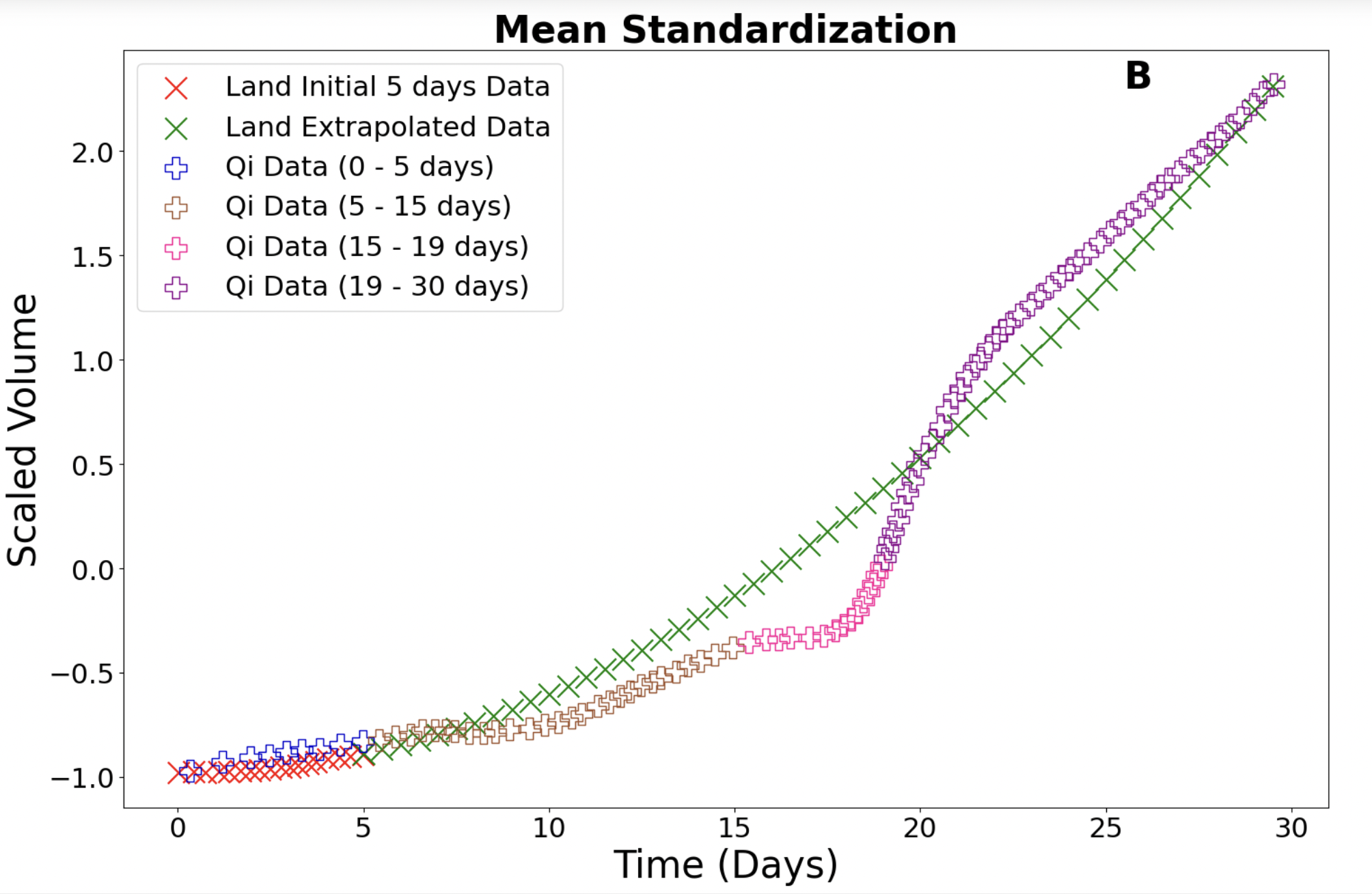}
    \caption{{\color{black}{Mean Standardization with respective shifting and scaling factors of 0.33 and 1.15, comparing Land, et al \cite{Land2019} versus Qi, et al \cite{qi2022} data.}}}
    \label{fig3b}
    \end{subfigure}
 \medskip
 \begin{subfigure}{0.49\textwidth}
     \includegraphics[width=\textwidth]{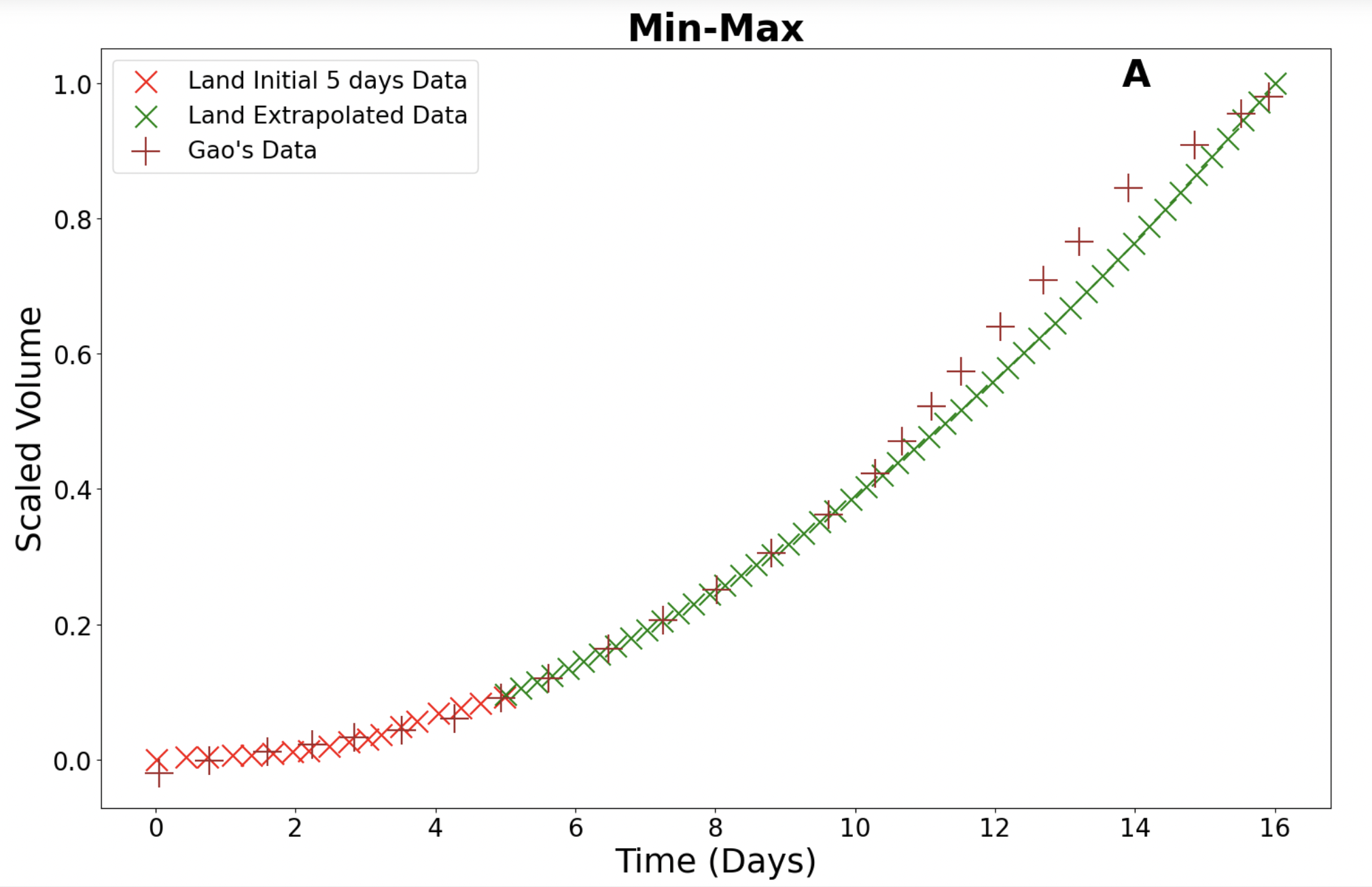}
    \caption{{\color{black}{Min-Max Standardization with respective shifting and scaling factors of 0.33 and 1.15, comparing Land, et al \cite{Land2019} versus Gao, et al \cite{gao2015} data.}}}
    \label{fig3c}
\end{subfigure}
 \hfill
 \begin{subfigure}{0.49\textwidth}
    \includegraphics[width=1\linewidth]{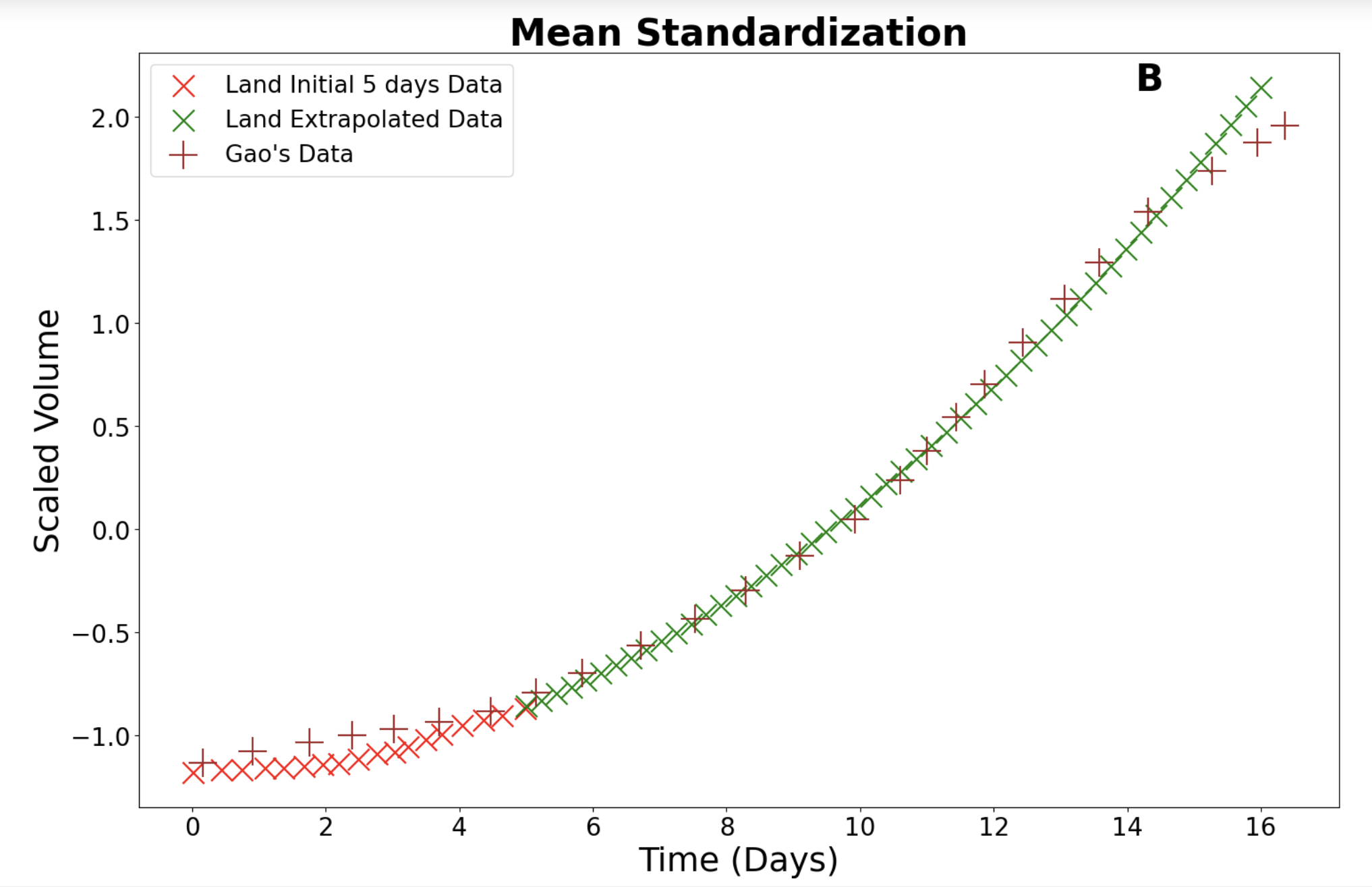}
    \caption{{\color{black}{Mean Standardization with respective shifting and scaling factors of 0.33 and 1.15, comparing Land, et al \cite{Land2019} versus Gao, et al \cite{gao2015} data.}}}
    \label{fig3d}
    \end{subfigure}
\caption{{\color{black}{Comparison of two top (best fit) standardization protocols from a choice of five standardization methods, trained on Land et al's initial 5-days  and extrapolated Saline data \cite{Land2019}(using Support Vector Regression (SVR)) over Qi et al and Gao et al's Saline Data \cite{gao2015,qi2022} respectively.}}}
 \label{fig3}
\end{figure}

\noindent
{\color{black}{The synthetic (Grabit-extracted) data from  Qi, et al \cite{qi2022} spanning 30 days is split into four segments to track different phases of tumor volume changes: \\\\
Segment 1: Days 0 to 5.\\
Segment 2: Days 5 to 15.\\
Segment 3: Days 15 to 19.\\
Segment 4: Days 19 to 30.\\

\noindent 
The Gao, et al data \cite{gao2015} is split into 2 segments to compare against similar segments from Land, et al \cite{Land2019} and Qi, et al \cite{qi2022}: \\\\
Segment 1: Days 0 to 5.\\
Segment 2: Days 5 to 16.\\
\noindent
AI models trained on Land, et al's saline data is probabilistically extrapolated over extended time periods of 25 and 11 days respectively using Support Vector Regression (SVR) to compare against studies by Qi, et al \cite{qi2022} (30 days data) and Gao, et al \cite{gao2015} (16 days data). Figures \ref{fig3} compare probabilistic fit of the Land, et al trained data against Qi, et al and the Gao, et al data respectively. Both mean and min-max protocols are used to project expected outcomes. The Min-Max scaling approach for both data sets, Qi, et al and Gao, et al, are adapted to identical shifting and scaling factors of 0.03 and 1.2 respectively, while Mean Standardization used similar factors of 0.33 and 1.15 respectively to ensure uniform baseline across the 3 different datasets. While SVR as a predictive model works well with the Gao, et al data (Figures \ref{fig3c} and \ref{fig3d}), where the Min-Max Standardization extrapolates the later predictive regimes even better than Mean Standardization, the predictions for Segments 2 and 3 for the Qi, et al data (Figures \ref{fig3a} and \ref{fig3b}) is clearly inefficient. This augurs the need for more powerful predictive tools.}} 

%

\begin{figure}[H]
 \centering
 \setkeys{Gin}{width=1.3\linewidth,height=0.35\textheight}
 \begin{subfigure}{0.49\textwidth}
     \includegraphics[width=\textwidth]{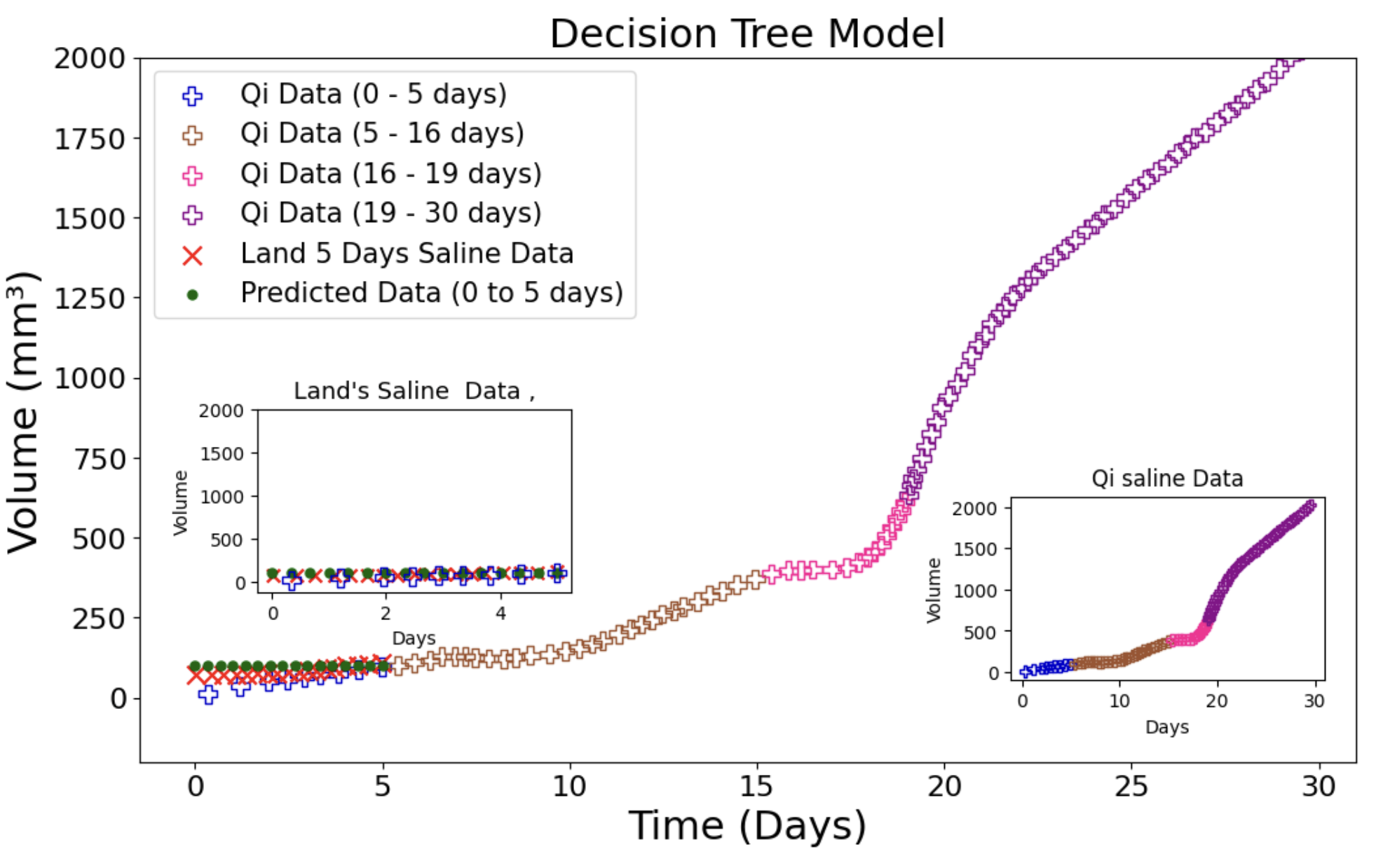}
     \caption{\bf Decision Tree Model}
     \label{fig4a}
 \end{subfigure}
 \hfill
 \begin{subfigure}{0.49\textwidth}
     \includegraphics[width=\textwidth]{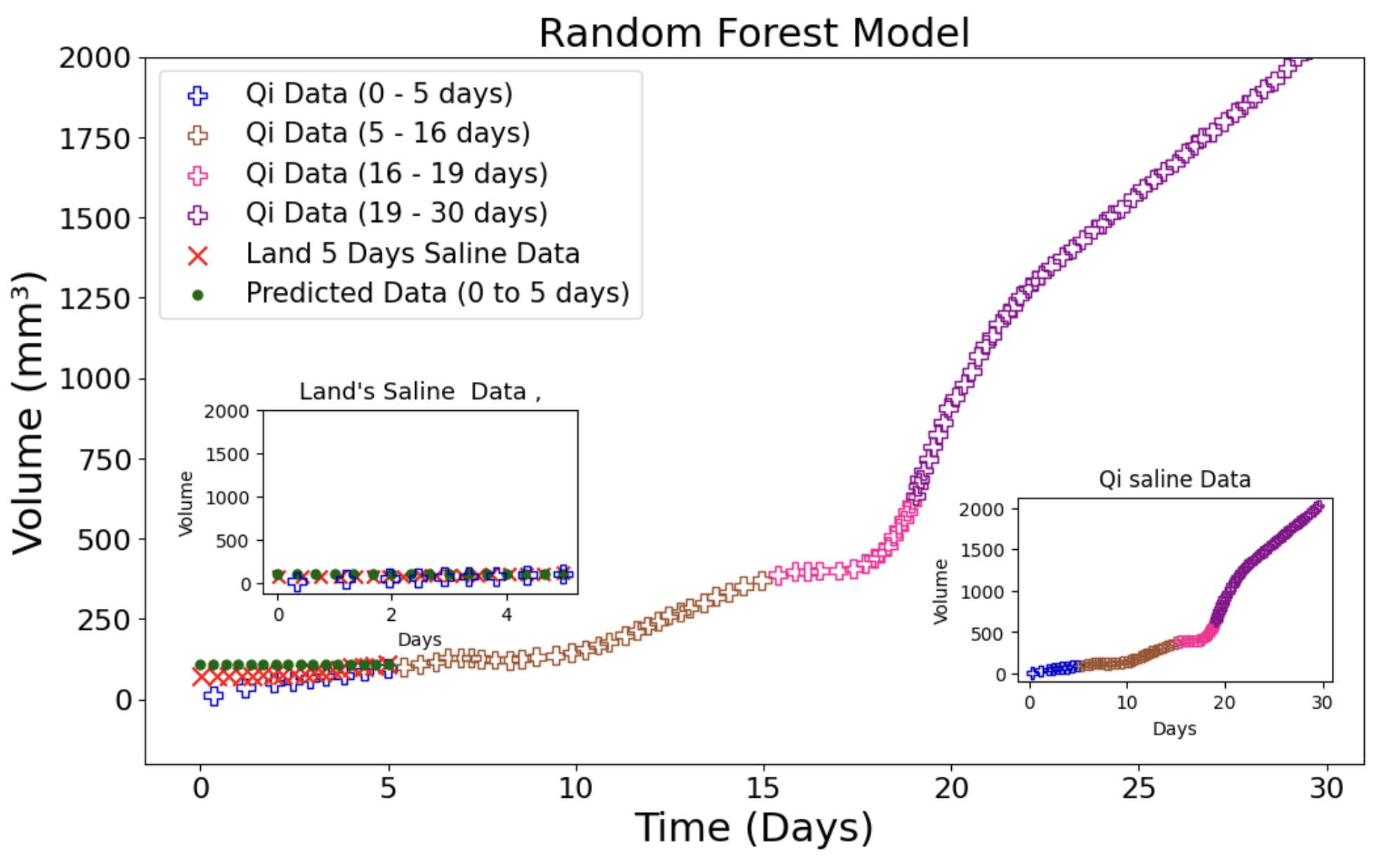}
     \caption{\bf Random Forest Model}
     \label{fig4b}
 \end{subfigure}
 
 \medskip
 \begin{subfigure}{0.49\textwidth}
     \includegraphics[width=\textwidth]{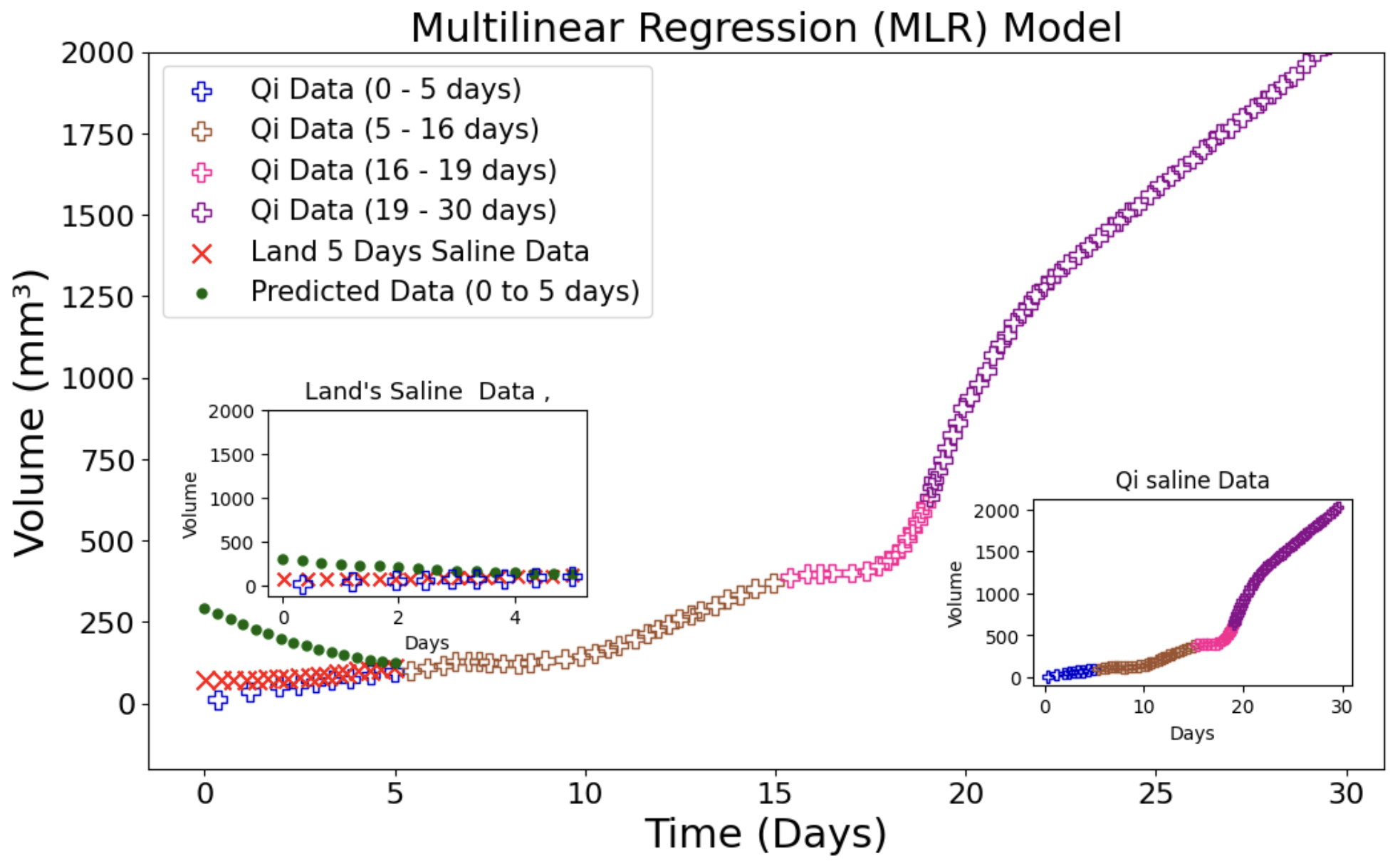}
     \caption{\bf Multilinear Regression (MLR)}
     \label{fig4c}
 \end{subfigure}
 \hfill
 \begin{subfigure}{0.49\textwidth}
     \includegraphics[width=\textwidth]{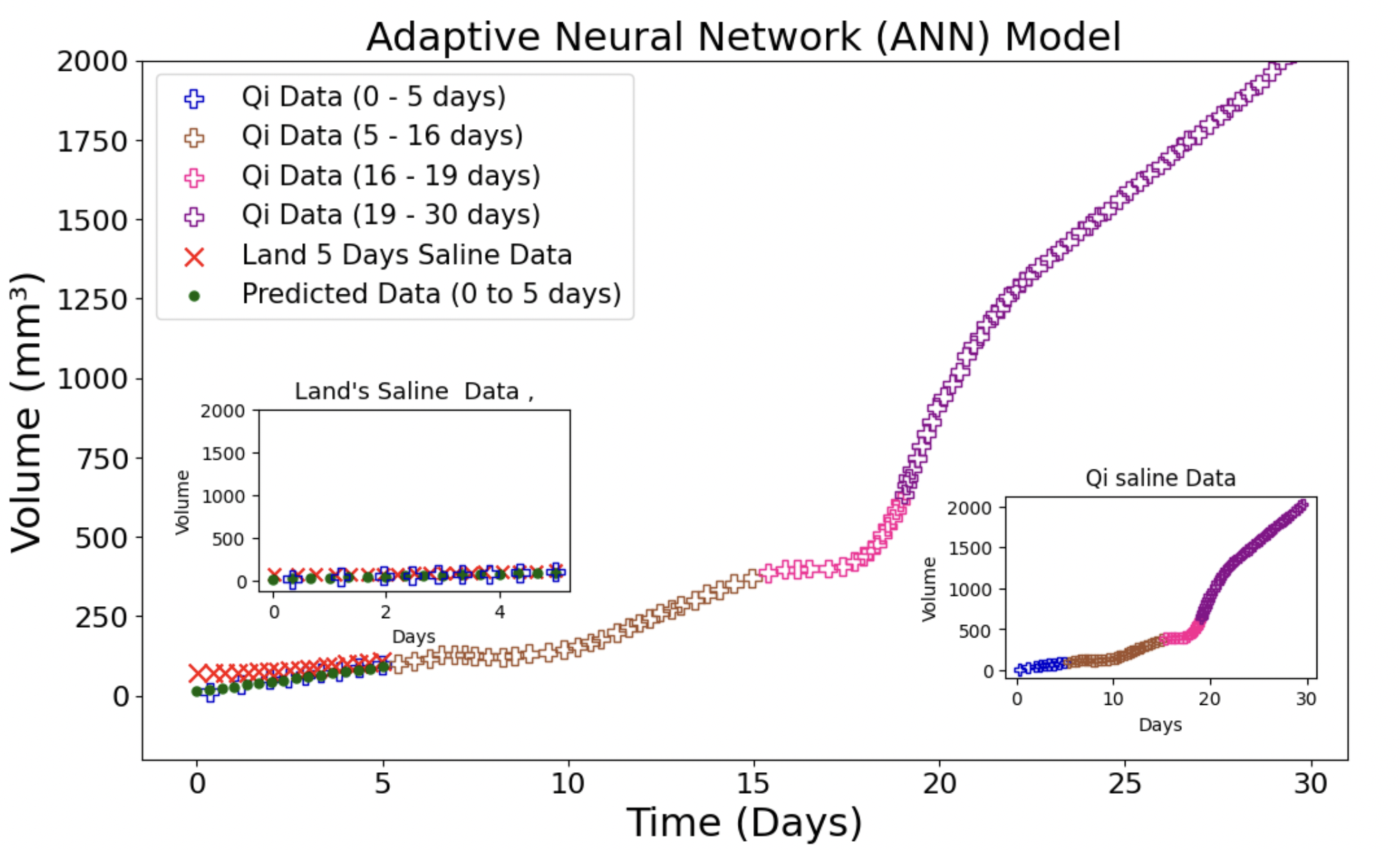}
     \caption{\bf Adaptive Neural Network (ANN)}
     \label{fig4d}
 \end{subfigure}
    \caption{First 5 days' predictions of Qi, et al's Saline \cite{qi2022} data against Land, et al's Saline data \cite{Land2019}, fitted with a scaling factor 0.4. The insets respectively demonstrate details of Land, et al versus Qi, et al salinity profiles for their respective times of study.}
    \label{fig4}
    \end{figure}

\begin{figure}[H]
 \centering
 \setkeys{Gin}{width=1.3\linewidth,height=0.35\textheight}
 \begin{subfigure}{0.49\textwidth}
     \includegraphics[width=\textwidth]{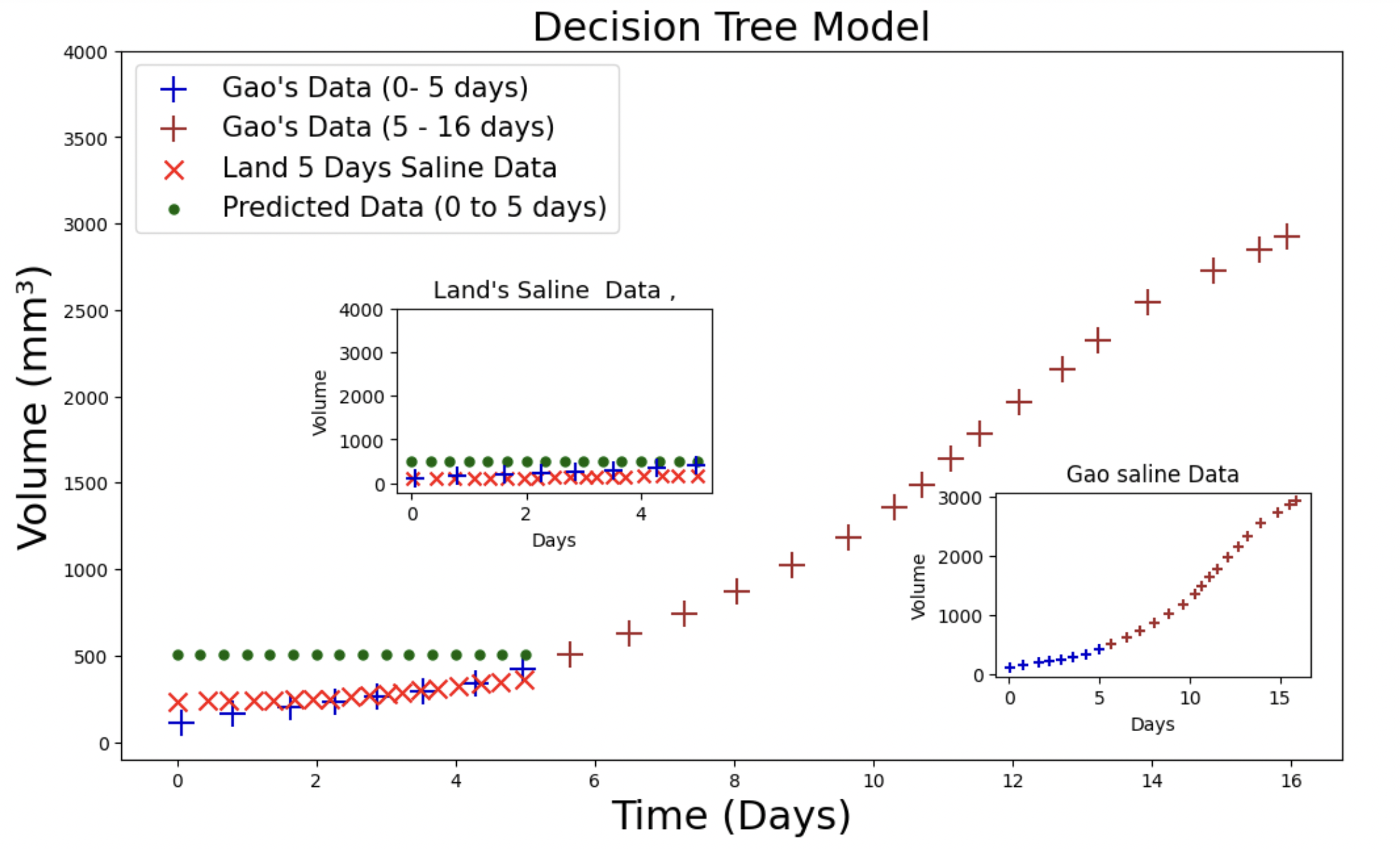}
     \caption{\bf Decision Tree Model}
     \label{fig5a}
 \end{subfigure}
 \hfill
 \begin{subfigure}{0.49\textwidth}
     \includegraphics[width=\textwidth]{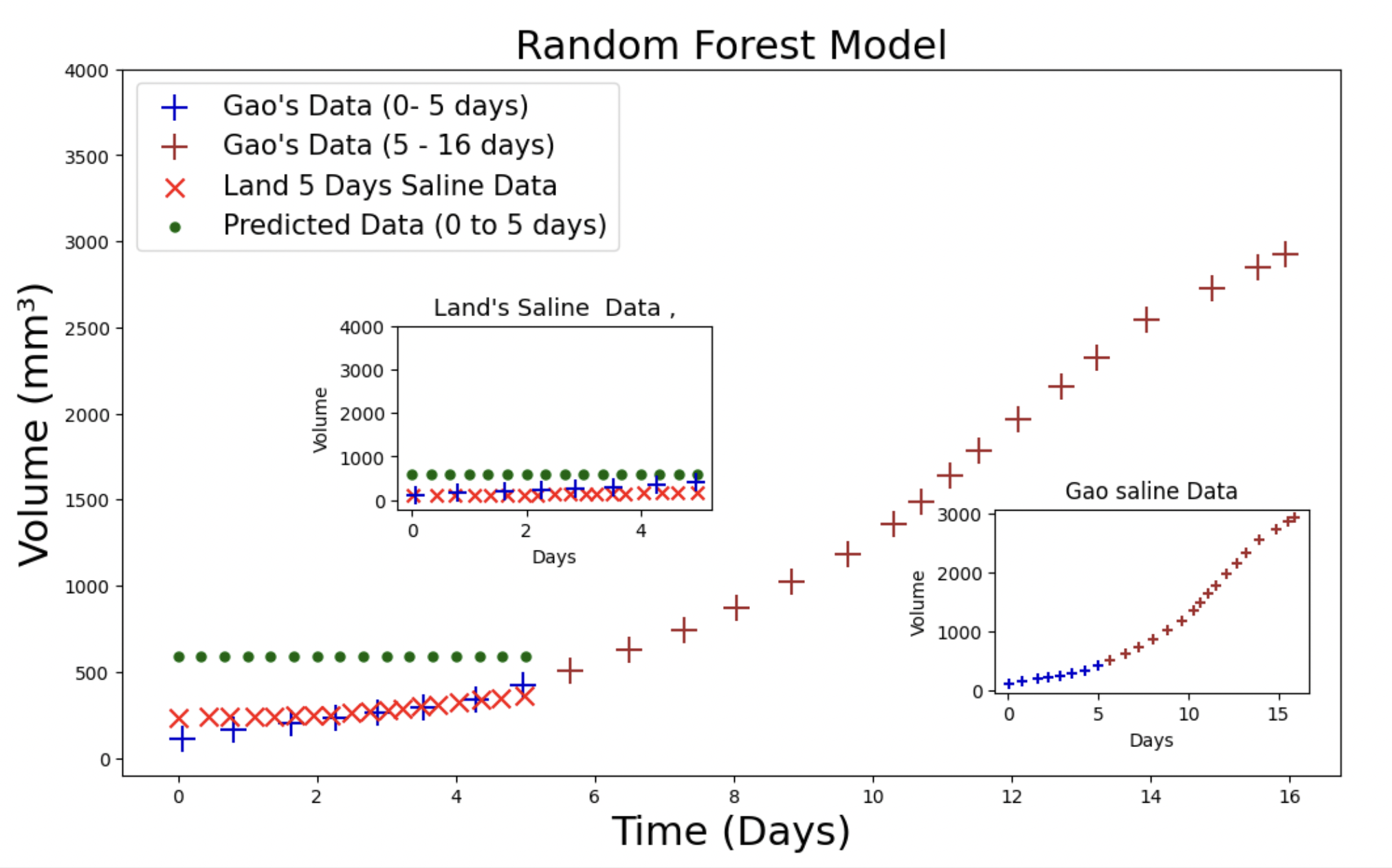}
     \caption{\bf Random Forest Model}
     \label{fig5b}
 \end{subfigure}
 
 \medskip
 \begin{subfigure}{0.49\textwidth}
     \includegraphics[width=\textwidth]{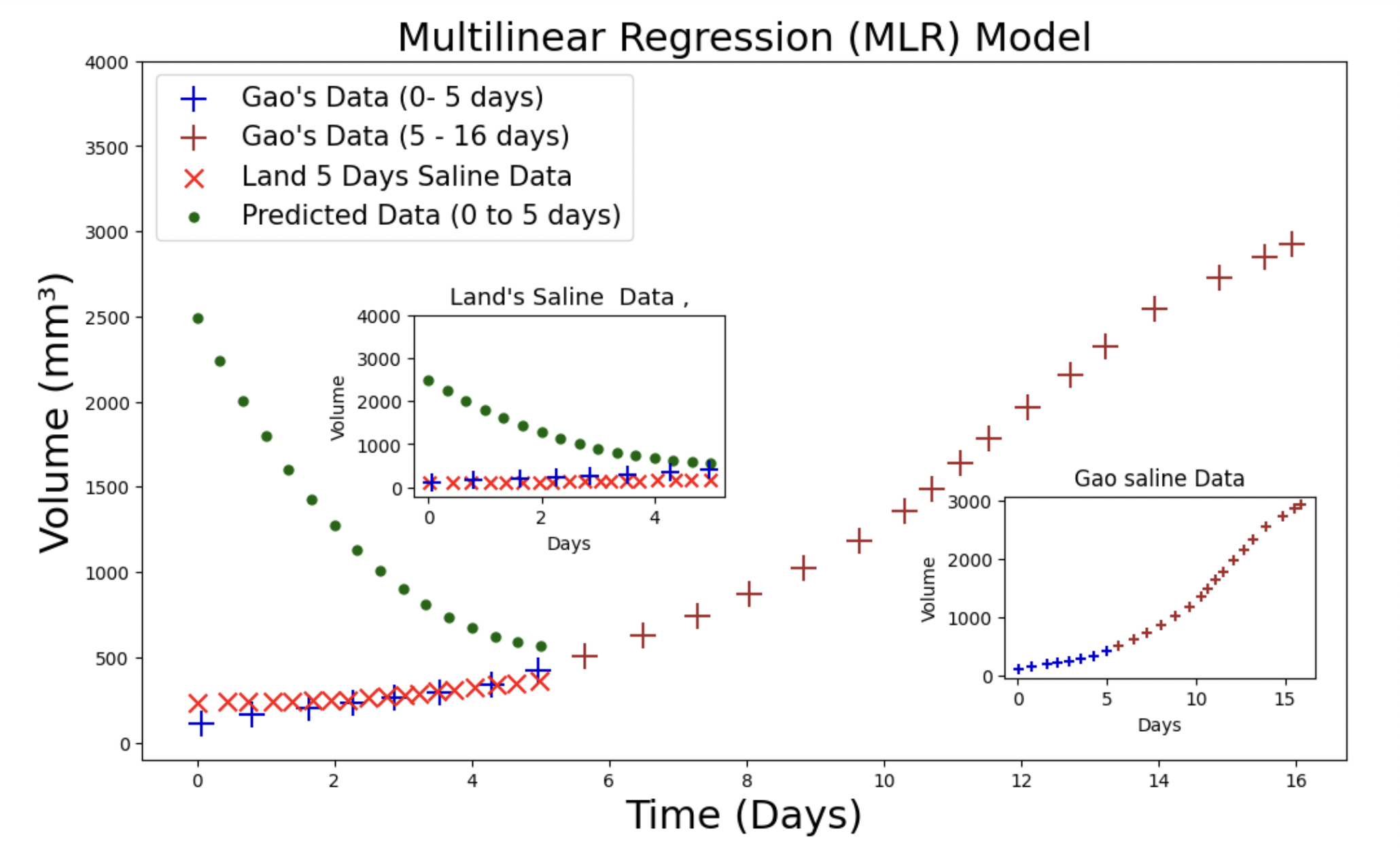}
     \caption{\bf Multilinear Regression (MLR)}
     \label{fig5c}
 \end{subfigure}
 \hfill
 \begin{subfigure}{0.49\textwidth}
     \includegraphics[width=\textwidth]{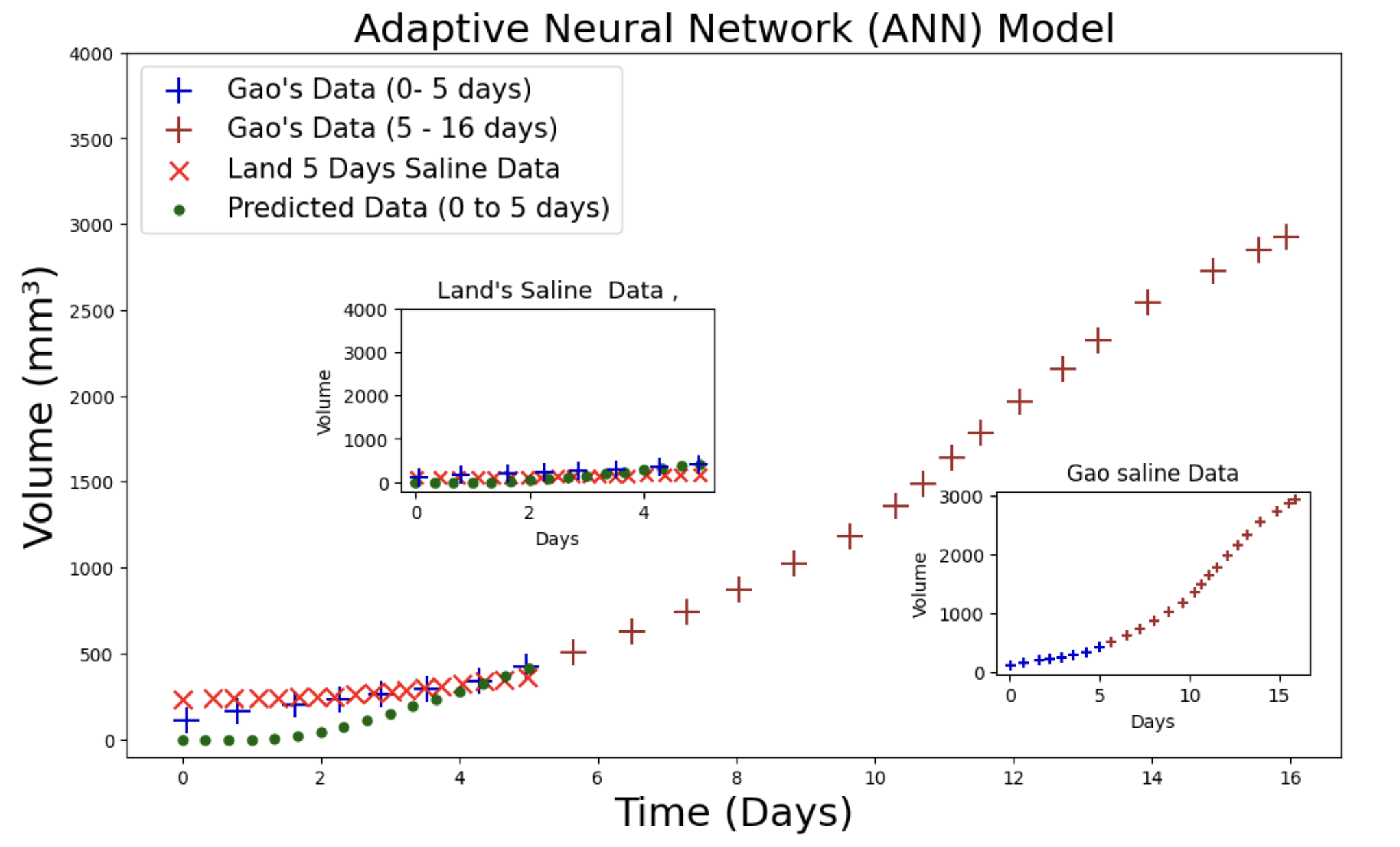}
     \caption{\bf Adaptive Neural Network (ANN)}
     \label{fig5d}
 \end{subfigure}
    \caption{First 5 days' predictions of Gao et al's Saline \cite{gao2015} data against Land et al's Saline data \cite{Land2019}, fitted with a scaling factor 0.55.The insets respectively demonstrate details of Land, et al versus Gao, et al salinity profiles for their respective times of study.}
    \label{fig5}
    \end{figure}
\begin{figure}[H]
  \setkeys{Gin}{width=1.2\linewidth,height=0.35\textheight}
 \begin{subfigure}{0.49\textwidth}
     \includegraphics[width=\textwidth]{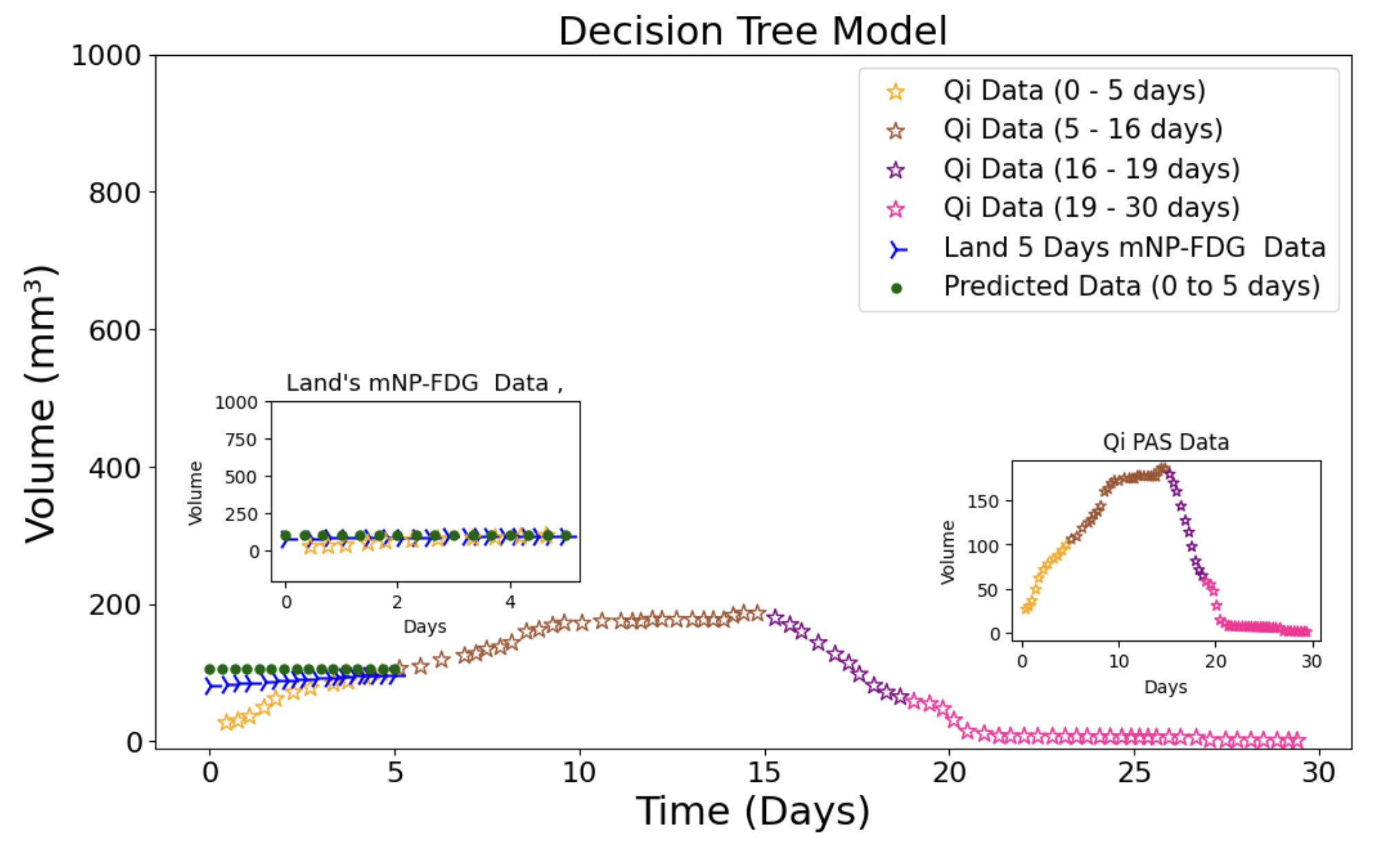}
     \caption{Decision Tree Model}
     \label{fig6a}
 \end{subfigure}
 \hfill
 \begin{subfigure}{0.49\textwidth}
     \includegraphics[width=\textwidth]{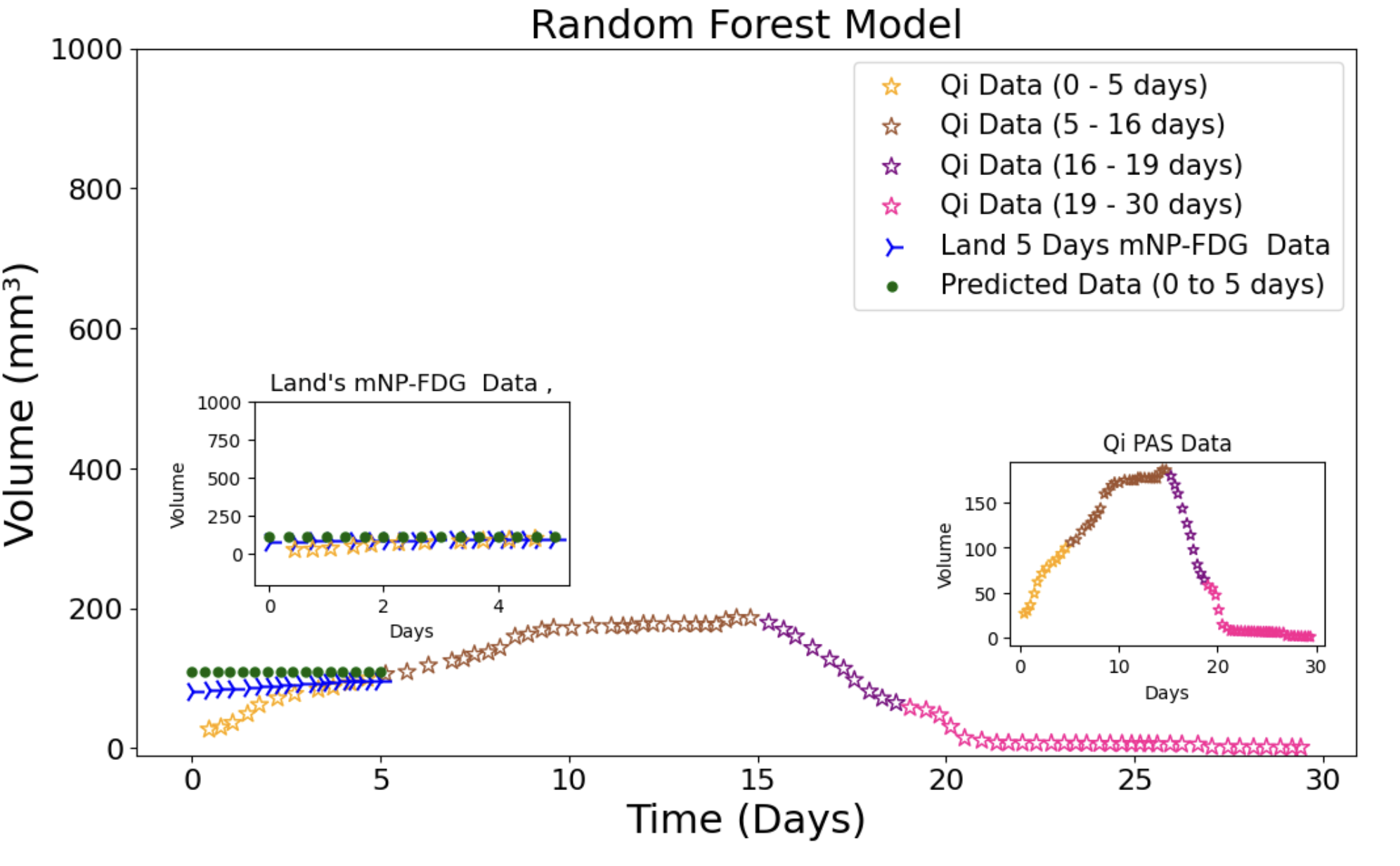}
     \caption{Random Forest Model}
     \label{fig6b}
 \end{subfigure}
 
 \medskip
 \begin{subfigure}{0.49\textwidth}
     \includegraphics[width=\textwidth]{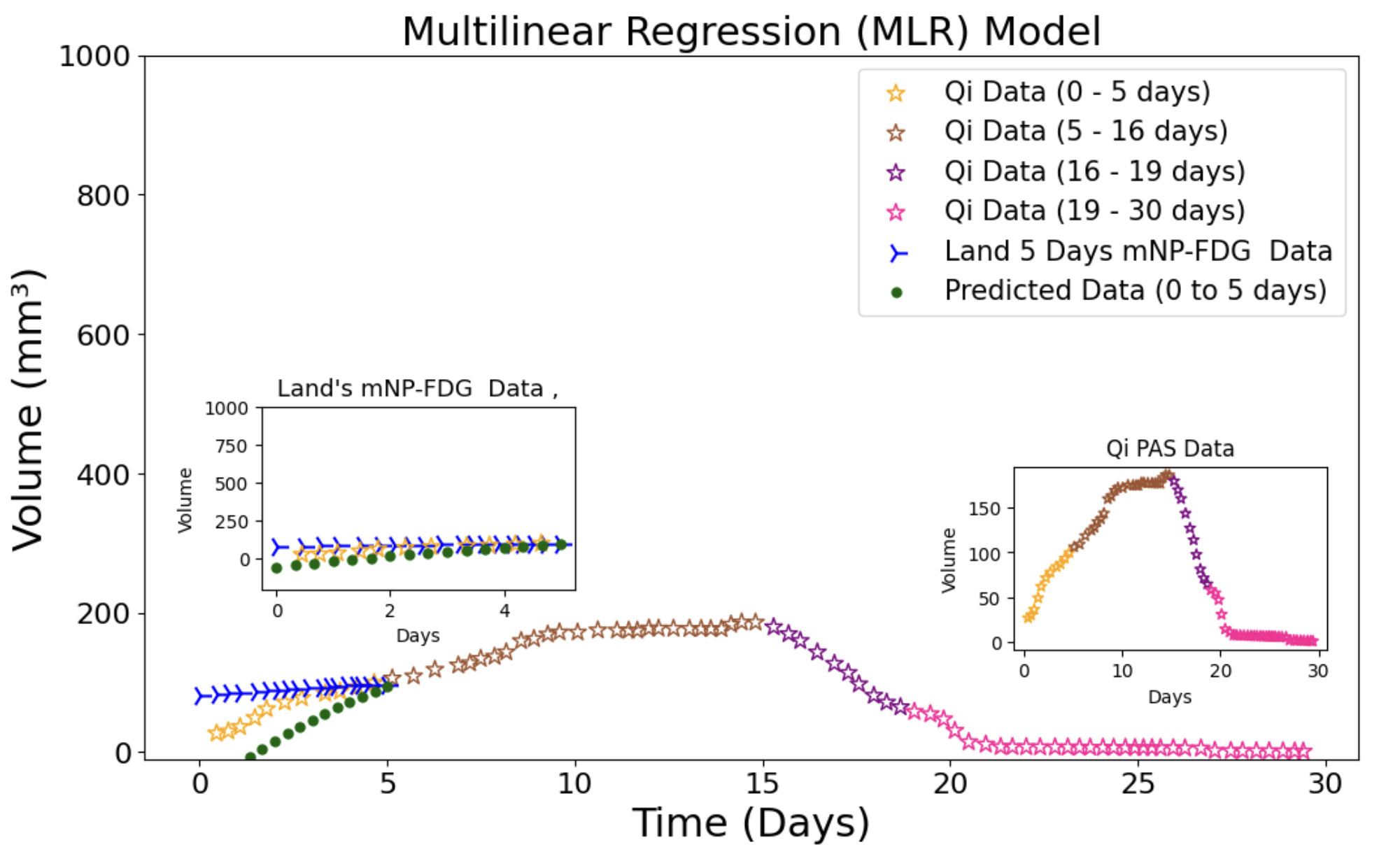}
     \caption{Multilinear Regression (MLR) Model}
     \label{fig6c}
 \end{subfigure}
 \hfill
 \begin{subfigure}{0.49\textwidth}
     \includegraphics[width=\textwidth]{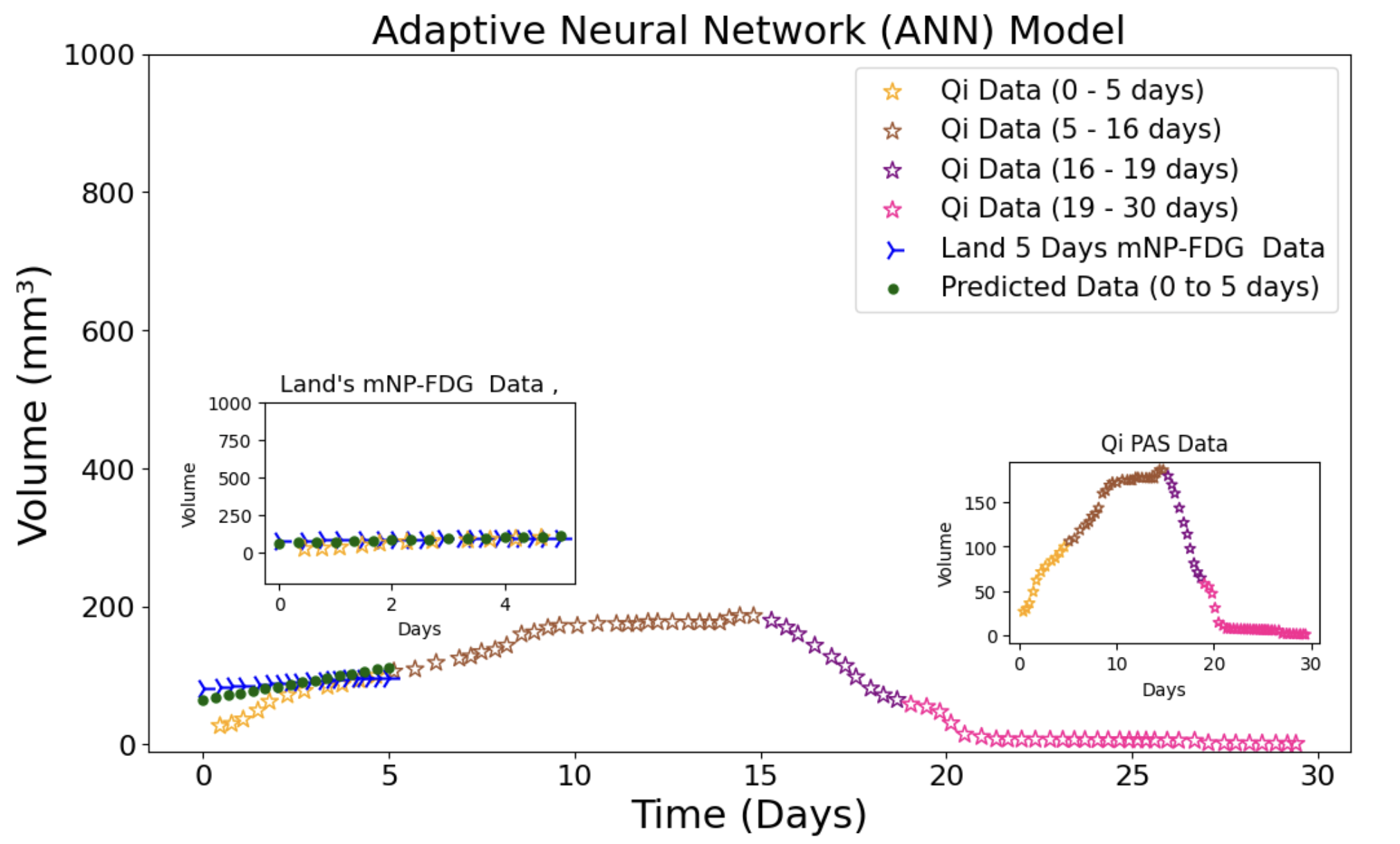}
     \caption{Adaptive Neural Network (ANN)}
     \label{fig6d}
 \end{subfigure}

    \caption{ First 5 days' predictions of Qi et al's PAS data \cite{qi2022} against Land et al's Saline \cite{Land2019} data fitted with a scaling factor 0.55. The insets respectively demonstrate details of Land, et al versus Qi, et al's mNP-FDG and PAS profiles for their respective times of study or predictions.}
    \label{fig6}

\end{figure}

For each segment, time and tumor volume data are divided and reshaped for model training. The training data from \enquote{Segment 2} is used to predict \enquote{Segment 1}. This approach is aimed to assess the accuracy and predictive capabilities of the models and evaluate the quality of agreement of Land 's data \cite{Land2019} when validated against the Qi's data \cite{qi2022}.
\noindent 
In Figure \ref{fig5} an \ref{fig6}, the top inset focuses on the data from Land, et al \cite{Land2019}, while the bottom inset shows Qi, et al and Gao, et al's data respectively \cite{qi2022,gao2015}, both under saline treatment. The data are analyzed using 3 Machine Learning (ML) models, the Decision Tree (DT), Random Forest (RF), and Multilinear Regression (MLR), alongside the Deep Learning Adaptive Neural Network (ANN) model to compare predictive capabilities. Clearly, the Multiple Linear Regression(MLR) model is the least satisfactory while the Deep Learning Adaptive Neural Network (ANN) method is the best predictive model amongst the methods used. 
\par 
The same models are then used to predict the expected performance of mNP-FDG as a cancer-containing agent Figure \ref{fig6}. Towards this, we fit the mNP-FDG model against Plasma Activated Saline (PAS) data because both treatments demonstrate tumor shrinkage. Our modelling predicts tumor shrinkage with mNP-FDG treatment in approximately 4.35 days compared to 9.58 days for the PAS solution. Qi et al's saline solution is not used here as it shows logistic and exponential growths (discussed under the 4 segments above).

\begin{figure}[H]
    \centering
    \includegraphics[width=0.9\linewidth]{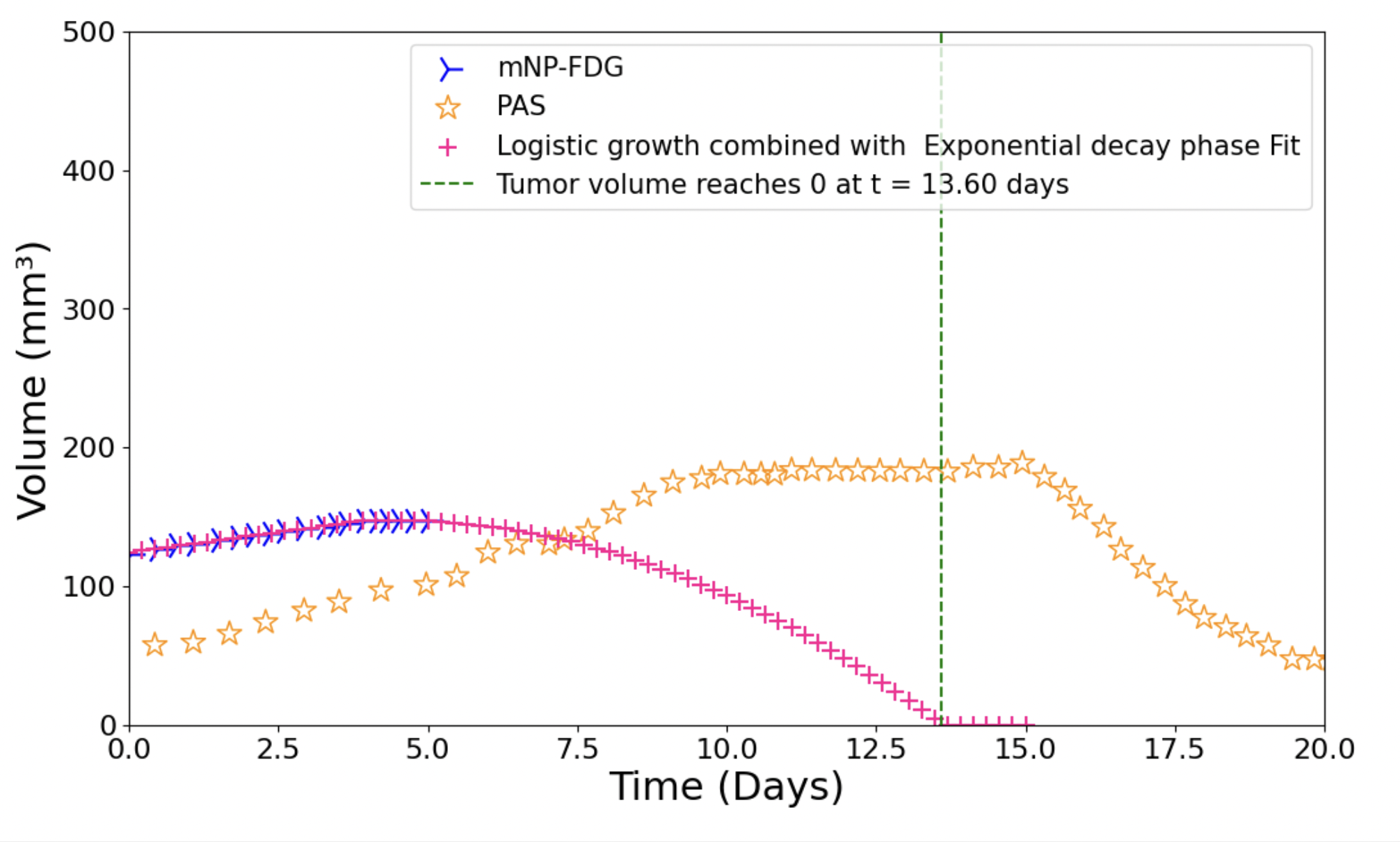}
    \caption{{\color{black}{Tumor Volume Predictions Using  Logistic Growth and Exponential Decay. At $t=13.60$ days, the tumor volume shrinks to zero (marked by a dotted line perpendicular to the time-axis). The prediction compares favorably with previous experimental evaluations reported in \cite{subramanian2016,Akin2018}.}}}
    \label{fig7}
\end{figure}

ANN is clearly the most consistent predictive model, a conclusion that could potentially help to determine  how the mNP-FDG treatment would perform in humans, guiding initial clinical trial doses through allometric scaling.  
PAS, characterised by both  growth and shrinkage phases, provides a foundation for predicting complete tumor shrinkage by mNP-FDG. Factors such as immune response, resistance rates and heterogeneity of the tumor are assumed minimal for the predictions. Comparing the rates of increase ($\text{PAS}_{\uparrow}$) and decrease ($\text{PAS}_{\downarrow}$) of PAS controlled tumor, we find $\text{PAS}_{\uparrow} = \frac{177.86 - 50}{9.58 - 0} = \frac{127.86}{9.58} \approx 13.34$ compared to $\text{PAS}_{\downarrow} = \frac{182.95 - 51}{12.90 - 28} = \frac{131.95}{-15.10} \approx -8.74$. Combining this with the statistics for the rate of increase of mNP-FDG in the body of mice ($\text{mNP-FDG}_{\uparrow}$), where $\text{mNP-FDG}_{\uparrow} = \frac{147.68 - 123.57}{4.52 - 0.013} = \frac{24.11}{4.507} \approx 5.34$, this allows us a rate kinetic estimation of the rate of tumor shrinkage due to mNP-FDG control $\text{mNP-FDG}_{\downarrow}$) as $\frac{13.34}{-8.74} = \frac{5.34}{\text{mNP-FDG}_{\downarrow}}$,
giving $\text{mNP-FDG}_{\downarrow} \approx -3.50$, a three-fold reduction in the tumor rate, that can be seen in Figure \ref{fig6}.
{\color{black}{To model the complex dynamics of tumor response as shown in Figure \ref{fig7}, we combine two functions, one based on faster logistic growth followed by relatively slower exponential decay.}} \\

\noindent    
{\it Logistic Growth Phase}:
\[
\text{logistic growth rate} = \frac{A_1}{1 + \exp(-k_1 \cdot (t - t_{\text{switch}}))}
\]

\noindent
{\it Exponential Decay Phase} (starting from \( t_{\text{switch}} \)):
\[
\text{exponential decay rate} = A_2 \cdot \exp(-k_2 \cdot (t - t_{\text{switch}})) - \alpha \cdot (t - t_{\text{switch}}) - \beta \cdot (t - t_{\text{switch}})^2
\]

\noindent
Here \( A_1 \) and \( A_2 \) are the respective amplitudes of the logistic growth and exponential decay phases, \( k_1 \) and \( k_2 \) the respective growth and decay rates, \( t \) is time, and \( t_{\text{switch}} \) is the time point at which the transition from logistic growth to exponential decay occurs. \( \alpha \) and \( \beta \) are coefficients that adjust the linear and quadratic terms in the decay phase.

 This model extrapolation successfully captured the initial phase of logistic growth followed by a gradual decline, mimicking the  tumor response to mNP-FDG treatment.

\noindent
 {\it Our model predicts total destruction of the tumor in ca 13.60 days after the initiation of treatment.} This is a key finding that awaits testing. These outcomes highlight mNP-FDG's potential as a promising treatment in oncology, capable of completely or partially reducing the tumor. 
\begin{table}[H]
\centering
\begin{tabular}{|c|c|c|}
\hline
\textbf{Treatment} & \textbf{Time (Days)} & \textbf{Reference} \\
\hline
Plasma Activated Saline (PAS) & 28 & \cite{gao2015} \\
\hline
mNP-FDG & 13  & {\bf First  results in this article} \\
\hline
Hypoxia-activated prodrugs (HAP) treatment & 14 & \cite{wilkins2023} \\
\hline
Exogenous Epidermal Growth  & 20 & \cite{lim2015}  \\
\hline
 GZD856& 16 & \cite{lu2017} \\
\hline
\end{tabular}
\caption{Significant Tumor Shrinkage}
\label{table1}
\end{table}

\section{Conclusion}
{\color{black}{
In this study, we explored and evaluated several models to predict cancerous tumor progression, comparing machine learning (Decision Tree, Random Forest, Multilinear Regression), a deep learning (Adaptive Neural Network(ANN) ) and {\color{black}  mathematical models (Logistic Growth and Exponential Decay)}.  Our results demonstrated that while the ANN model excelled in predicting tumor progression in saline-treated cases, it also demonstrated the superiority of mNP-FDG (fluorodeoxyglucose iron oxide magnetic nanoparticles) treatment. Our first key finding indicates that saline tumor treatment is best predicted using the ANN which was promising in its own right and set the stage to advance our research that shown mNP-FDG treatment as the best treatment regime demonstrating its potential as a transformative approach in oncology. This leads to the second key finding which is predicting a precise timeline for the cancer tumor to be entirely eradicated using mNP-FDG treatment. Using a combined Logistic Growth and Exponential Decay mode, that  successfully captures the initial growth followed by a gradual decline, this timeline is predicted at 13.60 days. This is a remarkable outcome that is open to laboratory verification. More importantly, the modeling architecture can deal with any other forms of treatment data to predict similar numbers to compare and find the best treatment regime for the specific form of cancer.

\par
The study demonstrates the power of artificial intelligence (AI) in predicting cancer tumor growth in mice subjected to treatments with saline and magnetic nano-particles. By leveraging the advanced computational capabilities of AI, we achieved highly accurate predictions, surpassing traditional machine learning methods in terms of precision and reliability. The ANN model effectively captures the complex, nonlinear relationships inherent in the biological data, allowing for nuanced and robust forecasting of tumor progression. This superiority underscores the potential of AI as a powerful tool in oncological research, offering a promising avenue for the development of more effective treatment strategies and personalized medicine.

\par
However, due to limited access to local data (only 5 days as in \cite{Land2019}), data from two established studies \cite{qi2022, gao2015}, pursuing similar treatment regimes, is used. The use of Artificial Intelligence (AI) methods provides unparalleled processing power to first validate and then predict tumor behaviors within the data from local (\cite{Land2019}) and external sources (\cite{qi2022, gao2015}).}} {\color{black}{We note that while the cancerous tumor under mNP-FDG treatment vanished in 13 days, similar treatment regimes using other agents as shown in Table \ref{table1} achieved the same goal of complete removal of the tumor. For all these alternative options, the time lines are much longer than 13.6 days though.}}

Logistic Growth and Exponential Decay model's effectiveness in predicting a 13.60-day timeline for mNP-FDG-induced complete tumor remission. This approach advances cancer treatment prediction, highlighting mNP-FDG as a superior treatment 
\par
Moreover, our comparative analysis reveals that while conventional machine learning techniques provide valuable insights, they are limited in handling the intricacies and high dimensionality of the dataset as efficiently as ANNs. The integration of magnetic nanoparticles in the treatment regimen further highlighted the model’s capacity to adapt to various therapeutic contexts and predict outcomes with significant accuracy. This research not only advances the application of ANN in cancer prognosis but also sets a precedent for future studies aiming to utilize AI in medical research. Consequently, the findings advocate for continued exploration and refinement of ANN methodologies to enhance their applicability and efficacy in clinical settings. Future work will focus on exploring additional cancer treatment agents using AI combined with allometric scaling, a technique that can potentially estimate and predict how mNP-FDG treatment can benefit cancer treatment in humans, guiding initial clinical trial dosage aiding non-invasive personalized cancer management using AI.

\section{Acknowledgments}
The authors acknowledge the postgraduate dissertation work by Megan Land as a source of some of the data used in this study. Also, Perihan Unak and Volkan Yaksakci for resourceful interactions and for contributing to studies on mNP-FDG.


\end{document}